
 File jytex.tex, for jyTeX version 2.6M (June 1992)
 Copyright (c) 1991, 1992 by Jonathan P. Yamron

\catcode`\@=11


\message{Loading jyTeX fonts...}



\font\vptrm=cmr5
\font\vptmit=cmmi5
\font\vptsy=cmsy5
\font\vptbf=cmbx5

\skewchar\vptmit='177 \skewchar\vptsy='60
\fontdimen16 \vptsy=\the\fontdimen17 \vptsy

\def\vpt{\ifmmode\err@badsizechange\else
     \@mathfontinit
     \textfont0=\vptrm  \scriptfont0=\vptrm  \scriptscriptfont0=\vptrm
     \textfont1=\vptmit \scriptfont1=\vptmit \scriptscriptfont1=\vptmit
     \textfont2=\vptsy  \scriptfont2=\vptsy  \scriptscriptfont2=\vptsy
     \textfont3=\xptex  \scriptfont3=\xptex  \scriptscriptfont3=\xptex
     \textfont\bffam=\vptbf
     \scriptfont\bffam=\vptbf
     \scriptscriptfont\bffam=\vptbf
     \@fontstyleinit
     \def\rm{\vptrm\fam=\z@}%
     \def\bf{\vptbf\fam=\bffam}%
     \def\oldstyle{\vptmit\fam=\@ne}%
     \rm\fi}


\font\viptrm=cmr6
\font\viptmit=cmmi6
\font\viptsy=cmsy6
\font\viptbf=cmbx6

\skewchar\viptmit='177 \skewchar\viptsy='60
\fontdimen16 \viptsy=\the\fontdimen17 \viptsy

\def\vipt{\ifmmode\err@badsizechange\else
     \@mathfontinit
     \textfont0=\viptrm  \scriptfont0=\vptrm  \scriptscriptfont0=\vptrm
     \textfont1=\viptmit \scriptfont1=\vptmit \scriptscriptfont1=\vptmit
     \textfont2=\viptsy  \scriptfont2=\vptsy  \scriptscriptfont2=\vptsy
     \textfont3=\xptex   \scriptfont3=\xptex  \scriptscriptfont3=\xptex
     \textfont\bffam=\viptbf
     \scriptfont\bffam=\vptbf
     \scriptscriptfont\bffam=\vptbf
     \@fontstyleinit
     \def\rm{\viptrm\fam=\z@}%
     \def\bf{\viptbf\fam=\bffam}%
     \def\oldstyle{\viptmit\fam=\@ne}%
     \rm\fi}


\font\viiptrm=cmr7
\font\viiptmit=cmmi7
\font\viiptsy=cmsy7
\font\viiptit=cmti7
\font\viiptbf=cmbx7

\skewchar\viiptmit='177 \skewchar\viiptsy='60
\fontdimen16 \viiptsy=\the\fontdimen17 \viiptsy

\def\viipt{\ifmmode\err@badsizechange\else
     \@mathfontinit
     \textfont0=\viiptrm  \scriptfont0=\vptrm  \scriptscriptfont0=\vptrm
     \textfont1=\viiptmit \scriptfont1=\vptmit \scriptscriptfont1=\vptmit
     \textfont2=\viiptsy  \scriptfont2=\vptsy  \scriptscriptfont2=\vptsy
     \textfont3=\xptex    \scriptfont3=\xptex  \scriptscriptfont3=\xptex
     \textfont\itfam=\viiptit
     \scriptfont\itfam=\viiptit
     \scriptscriptfont\itfam=\viiptit
     \textfont\bffam=\viiptbf
     \scriptfont\bffam=\vptbf
     \scriptscriptfont\bffam=\vptbf
     \@fontstyleinit
     \def\rm{\viiptrm\fam=\z@}%
     \def\it{\viiptit\fam=\itfam}%
     \def\bf{\viiptbf\fam=\bffam}%
     \def\oldstyle{\viiptmit\fam=\@ne}%
     \rm\fi}


\font\viiiptrm=cmr8
\font\viiiptmit=cmmi8
\font\viiiptsy=cmsy8
\font\viiiptit=cmti8
\font\viiiptbf=cmbx8

\skewchar\viiiptmit='177 \skewchar\viiiptsy='60
\fontdimen16 \viiiptsy=\the\fontdimen17 \viiiptsy

\def\viiipt{\ifmmode\err@badsizechange\else
     \@mathfontinit
     \textfont0=\viiiptrm  \scriptfont0=\viptrm  \scriptscriptfont0=\vptrm
     \textfont1=\viiiptmit \scriptfont1=\viptmit \scriptscriptfont1=\vptmit
     \textfont2=\viiiptsy  \scriptfont2=\viptsy  \scriptscriptfont2=\vptsy
     \textfont3=\xptex     \scriptfont3=\xptex   \scriptscriptfont3=\xptex
     \textfont\itfam=\viiiptit
     \scriptfont\itfam=\viiptit
     \scriptscriptfont\itfam=\viiptit
     \textfont\bffam=\viiiptbf
     \scriptfont\bffam=\viptbf
     \scriptscriptfont\bffam=\vptbf
     \@fontstyleinit
     \def\rm{\viiiptrm\fam=\z@}%
     \def\it{\viiiptit\fam=\itfam}%
     \def\bf{\viiiptbf\fam=\bffam}%
     \def\oldstyle{\viiiptmit\fam=\@ne}%
     \rm\fi}


\def\getixpt{%
     \font\ixptrm=cmr9
     \font\ixptmit=cmmi9
     \font\ixptsy=cmsy9
     \font\ixptit=cmti9
     \font\ixptbf=cmbx9
     \skewchar\ixptmit='177 \skewchar\ixptsy='60
     \fontdimen16 \ixptsy=\the\fontdimen17 \ixptsy}

\def\ixpt{\ifmmode\err@badsizechange\else
     \@mathfontinit
     \textfont0=\ixptrm  \scriptfont0=\viiptrm  \scriptscriptfont0=\vptrm
     \textfont1=\ixptmit \scriptfont1=\viiptmit \scriptscriptfont1=\vptmit
     \textfont2=\ixptsy  \scriptfont2=\viiptsy  \scriptscriptfont2=\vptsy
     \textfont3=\xptex   \scriptfont3=\xptex    \scriptscriptfont3=\xptex
     \textfont\itfam=\ixptit
     \scriptfont\itfam=\viiptit
     \scriptscriptfont\itfam=\viiptit
     \textfont\bffam=\ixptbf
     \scriptfont\bffam=\viiptbf
     \scriptscriptfont\bffam=\vptbf
     \@fontstyleinit
     \def\rm{\ixptrm\fam=\z@}%
     \def\it{\ixptit\fam=\itfam}%
     \def\bf{\ixptbf\fam=\bffam}%
     \def\oldstyle{\ixptmit\fam=\@ne}%
     \rm\fi}


\font\xptrm=cmr10
\font\xptmit=cmmi10
\font\xptsy=cmsy10
\font\xptex=cmex10
\font\xptit=cmti10
\font\xptsl=cmsl10
\font\xptbf=cmbx10
\font\xpttt=cmtt10
\font\xptss=cmss10
\font\xptsc=cmcsc10
\font\xptbfs=cmbx10
\font\xptbmit=cmmib10

\skewchar\xptmit='177 \skewchar\xptbmit='177 \skewchar\xptsy='60
\fontdimen16 \xptsy=\the\fontdimen17 \xptsy

\def\xpt{\ifmmode\err@badsizechange\else
     \@mathfontinit
     \textfont0=\xptrm  \scriptfont0=\viiptrm  \scriptscriptfont0=\vptrm
     \textfont1=\xptmit \scriptfont1=\viiptmit \scriptscriptfont1=\vptmit
     \textfont2=\xptsy  \scriptfont2=\viiptsy  \scriptscriptfont2=\vptsy
     \textfont3=\xptex  \scriptfont3=\xptex    \scriptscriptfont3=\xptex
     \textfont\itfam=\xptit
     \scriptfont\itfam=\viiptit
     \scriptscriptfont\itfam=\viiptit
     \textfont\bffam=\xptbf
     \scriptfont\bffam=\viiptbf
     \scriptscriptfont\bffam=\vptbf
     \textfont\bfsfam=\xptbfs
     \scriptfont\bfsfam=\viiptbf
     \scriptscriptfont\bfsfam=\vptbf
     \textfont\bmitfam=\xptbmit
     \scriptfont\bmitfam=\viiptmit
     \scriptscriptfont\bmitfam=\vptmit
     \@fontstyleinit
     \def\rm{\xptrm\fam=\z@}%
     \def\it{\xptit\fam=\itfam}%
     \def\sl{\xptsl}%
     \def\bf{\xptbf\fam=\bffam}%
     \def\tt{\xpttt}%
     \def\ss{\xptss}%
     \def\sc{\xptsc}%
     \def\bfs{\xptbfs\fam=\bfsfam}%
     \def\bmit{\fam=\bmitfam}%
     \def\oldstyle{\xptmit\fam=\@ne}%
     \rm\fi}


\def\getxipt{
     \font\xiptrm=cmr10  scaled\magstephalf
     \font\xiptmit=cmmi10 scaled\magstephalf
     \font\xiptsy=cmsy10 scaled\magstephalf
     \font\xiptex=cmex10 scaled\magstephalf
     \font\xiptit=cmti10 scaled\magstephalf
     \font\xiptsl=cmsl10 scaled\magstephalf
     \font\xiptbf=cmbx10 scaled\magstephalf
     \font\xipttt=cmtt10 scaled\magstephalf
     \font\xiptss=cmss10 scaled\magstephalf
     \font\xiptsc=cmcsc10 scaled\magstephalf
     \skewchar\xiptmit='177 \skewchar\xiptsy='60
     \fontdimen16 \xiptsy=\the\fontdimen17 \xiptsy}

\def\xipt{\ifmmode\err@badsizechange\else
     \@mathfontinit
     \textfont0=\xiptrm  \scriptfont0=\viiiptrm  \scriptscriptfont0=\viptrm
     \textfont1=\xiptmit \scriptfont1=\viiiptmit \scriptscriptfont1=\viptmit
     \textfont2=\xiptsy  \scriptfont2=\viiiptsy  \scriptscriptfont2=\viptsy
     \textfont3=\xiptex  \scriptfont3=\xptex     \scriptscriptfont3=\xptex
     \textfont\itfam=\xiptit
     \scriptfont\itfam=\viiiptit
     \scriptscriptfont\itfam=\viiptit
     \textfont\bffam=\xiptbf
     \scriptfont\bffam=\viiiptbf
     \scriptscriptfont\bffam=\viptbf
     \@fontstyleinit
     \def\rm{\xiptrm\fam=\z@}
     \def\it{\xiptit\fam=\itfam}
     \def\sl{\xiptsl}
     \def\bf{\xiptbf\fam=\bffam}
     \def\tt{\xipttt}
     \def\ss{\xiptss}
     \def\sc{\xiptsc}
     \def\oldstyle{\xiptmit\fam=\@ne}
     \rm\fi}


\font\xiiptrm=cmr12
\font\xiiptmit=cmmi12
\font\xiiptsy=cmsy10  scaled\magstep1
\font\xiiptex=cmex10  scaled\magstep1
\font\xiiptit=cmti12
\font\xiiptsl=cmsl12
\font\xiiptbf=cmbx12
\font\xiipttt=cmtt12
\font\xiiptss=cmss12
\font\xiiptsc=cmcsc10 scaled\magstep1
\font\xiiptbfs=cmbx10  scaled\magstep1
\font\xiiptbmit=cmmib10 scaled\magstep1

\skewchar\xiiptmit='177 \skewchar\xiiptbmit='177 \skewchar\xiiptsy='60
\fontdimen16 \xiiptsy=\the\fontdimen17 \xiiptsy

\def\xiipt{\ifmmode\err@badsizechange\else
     \@mathfontinit
     \textfont0=\xiiptrm  \scriptfont0=\viiiptrm  \scriptscriptfont0=\viptrm
     \textfont1=\xiiptmit \scriptfont1=\viiiptmit \scriptscriptfont1=\viptmit
     \textfont2=\xiiptsy  \scriptfont2=\viiiptsy  \scriptscriptfont2=\viptsy
     \textfont3=\xiiptex  \scriptfont3=\xptex     \scriptscriptfont3=\xptex
     \textfont\itfam=\xiiptit
     \scriptfont\itfam=\viiiptit
     \scriptscriptfont\itfam=\viiptit
     \textfont\bffam=\xiiptbf
     \scriptfont\bffam=\viiiptbf
     \scriptscriptfont\bffam=\viptbf
     \textfont\bfsfam=\xiiptbfs
     \scriptfont\bfsfam=\viiiptbf
     \scriptscriptfont\bfsfam=\viptbf
     \textfont\bmitfam=\xiiptbmit
     \scriptfont\bmitfam=\viiiptmit
     \scriptscriptfont\bmitfam=\viptmit
     \@fontstyleinit
     \def\rm{\xiiptrm\fam=\z@}%
     \def\it{\xiiptit\fam=\itfam}%
     \def\sl{\xiiptsl}%
     \def\bf{\xiiptbf\fam=\bffam}%
     \def\tt{\xiipttt}%
     \def\ss{\xiiptss}%
     \def\sc{\xiiptsc}%
     \def\bfs{\xiiptbfs\fam=\bfsfam}%
     \def\bmit{\fam=\bmitfam}%
     \def\oldstyle{\xiiptmit\fam=\@ne}%
     \rm\fi}


\def\getxiiipt{%
     \font\xiiiptrm=cmr12  scaled\magstephalf
     \font\xiiiptmit=cmmi12 scaled\magstephalf
     \font\xiiiptsy=cmsy9  scaled\magstep2
     \font\xiiiptit=cmti12 scaled\magstephalf
     \font\xiiiptsl=cmsl12 scaled\magstephalf
     \font\xiiiptbf=cmbx12 scaled\magstephalf
     \font\xiiipttt=cmtt12 scaled\magstephalf
     \font\xiiiptss=cmss12 scaled\magstephalf
     \skewchar\xiiiptmit='177 \skewchar\xiiiptsy='60
     \fontdimen16 \xiiiptsy=\the\fontdimen17 \xiiiptsy}

\def\xiiipt{\ifmmode\err@badsizechange\else
     \@mathfontinit
     \textfont0=\xiiiptrm  \scriptfont0=\xptrm  \scriptscriptfont0=\viiptrm
     \textfont1=\xiiiptmit \scriptfont1=\xptmit \scriptscriptfont1=\viiptmit
     \textfont2=\xiiiptsy  \scriptfont2=\xptsy  \scriptscriptfont2=\viiptsy
     \textfont3=\xivptex   \scriptfont3=\xptex  \scriptscriptfont3=\xptex
     \textfont\itfam=\xiiiptit
     \scriptfont\itfam=\xptit
     \scriptscriptfont\itfam=\viiptit
     \textfont\bffam=\xiiiptbf
     \scriptfont\bffam=\xptbf
     \scriptscriptfont\bffam=\viiptbf
     \@fontstyleinit
     \def\rm{\xiiiptrm\fam=\z@}%
     \def\it{\xiiiptit\fam=\itfam}%
     \def\sl{\xiiiptsl}%
     \def\bf{\xiiiptbf\fam=\bffam}%
     \def\tt{\xiiipttt}%
     \def\ss{\xiiiptss}%
     \def\oldstyle{\xiiiptmit\fam=\@ne}%
     \rm\fi}


\font\xivptrm=cmr12   scaled\magstep1
\font\xivptmit=cmmi12  scaled\magstep1
\font\xivptsy=cmsy10  scaled\magstep2
\font\xivptex=cmex10  scaled\magstep2
\font\xivptit=cmti12  scaled\magstep1
\font\xivptsl=cmsl12  scaled\magstep1
\font\xivptbf=cmbx12  scaled\magstep1
\font\xivpttt=cmtt12  scaled\magstep1
\font\xivptss=cmss12  scaled\magstep1
\font\xivptsc=cmcsc10 scaled\magstep2
\font\xivptbfs=cmbx10  scaled\magstep2
\font\xivptbmit=cmmib10 scaled\magstep2

\skewchar\xivptmit='177 \skewchar\xivptbmit='177 \skewchar\xivptsy='60
\fontdimen16 \xivptsy=\the\fontdimen17 \xivptsy

\def\xivpt{\ifmmode\err@badsizechange\else
     \@mathfontinit
     \textfont0=\xivptrm  \scriptfont0=\xptrm  \scriptscriptfont0=\viiptrm
     \textfont1=\xivptmit \scriptfont1=\xptmit \scriptscriptfont1=\viiptmit
     \textfont2=\xivptsy  \scriptfont2=\xptsy  \scriptscriptfont2=\viiptsy
     \textfont3=\xivptex  \scriptfont3=\xptex  \scriptscriptfont3=\xptex
     \textfont\itfam=\xivptit
     \scriptfont\itfam=\xptit
     \scriptscriptfont\itfam=\viiptit
     \textfont\bffam=\xivptbf
     \scriptfont\bffam=\xptbf
     \scriptscriptfont\bffam=\viiptbf
     \textfont\bfsfam=\xivptbfs
     \scriptfont\bfsfam=\xptbfs
     \scriptscriptfont\bfsfam=\viiptbf
     \textfont\bmitfam=\xivptbmit
     \scriptfont\bmitfam=\xptbmit
     \scriptscriptfont\bmitfam=\viiptmit
     \@fontstyleinit
     \def\rm{\xivptrm\fam=\z@}%
     \def\it{\xivptit\fam=\itfam}%
     \def\sl{\xivptsl}%
     \def\bf{\xivptbf\fam=\bffam}%
     \def\tt{\xivpttt}%
     \def\ss{\xivptss}%
     \def\sc{\xivptsc}%
     \def\bfs{\xivptbfs\fam=\bfsfam}%
     \def\bmit{\fam=\bmitfam}%
     \def\oldstyle{\xivptmit\fam=\@ne}%
     \rm\fi}


\font\xviiptrm=cmr17
\font\xviiptmit=cmmi12 scaled\magstep2
\font\xviiptsy=cmsy10 scaled\magstep3
\font\xviiptex=cmex10 scaled\magstep3
\font\xviiptit=cmti12 scaled\magstep2
\font\xviiptbf=cmbx12 scaled\magstep2
\font\xviiptbfs=cmbx10 scaled\magstep3

\skewchar\xviiptmit='177 \skewchar\xviiptsy='60
\fontdimen16 \xviiptsy=\the\fontdimen17 \xviiptsy

\def\xviipt{\ifmmode\err@badsizechange\else
     \@mathfontinit
     \textfont0=\xviiptrm  \scriptfont0=\xiiptrm  \scriptscriptfont0=\viiiptrm
     \textfont1=\xviiptmit \scriptfont1=\xiiptmit \scriptscriptfont1=\viiiptmit
     \textfont2=\xviiptsy  \scriptfont2=\xiiptsy  \scriptscriptfont2=\viiiptsy
     \textfont3=\xviiptex  \scriptfont3=\xiiptex  \scriptscriptfont3=\xptex
     \textfont\itfam=\xviiptit
     \scriptfont\itfam=\xiiptit
     \scriptscriptfont\itfam=\viiiptit
     \textfont\bffam=\xviiptbf
     \scriptfont\bffam=\xiiptbf
     \scriptscriptfont\bffam=\viiiptbf
     \textfont\bfsfam=\xviiptbfs
     \scriptfont\bfsfam=\xiiptbfs
     \scriptscriptfont\bfsfam=\viiiptbf
     \@fontstyleinit
     \def\rm{\xviiptrm\fam=\z@}%
     \def\it{\xviiptit\fam=\itfam}%
     \def\bf{\xviiptbf\fam=\bffam}%
     \def\bfs{\xviiptbfs\fam=\bfsfam}%
     \def\oldstyle{\xviiptmit\fam=\@ne}%
     \rm\fi}


\font\xxiptrm=cmr17  scaled\magstep1


\def\xxipt{\ifmmode\err@badsizechange\else
     \@mathfontinit
     \@fontstyleinit
     \def\rm{\xxiptrm\fam=\z@}%
     \rm\fi}


\font\xxvptrm=cmr17  scaled\magstep2


\def\xxvpt{\ifmmode\err@badsizechange\else
     \@mathfontinit
     \@fontstyleinit
     \def\rm{\xxvptrm\fam=\z@}%
     \rm\fi}




\message{Loading jyTeX macros...}

\message{modifications to plain.tex,}


\def\newcount{\alloc@0\count\countdef\insc@unt}
\def\newdimen{\alloc@1\dimen\dimendef\insc@unt}
\def\newskip{\alloc@2\skip\skipdef\insc@unt}
\def\newmuskip{\alloc@3\muskip\muskipdef\@cclvi}
\def\newbox{\alloc@4\box\chardef\insc@unt}
\def\newtoks{\alloc@5\toks\toksdef\@cclvi}
\def\newhelp#1#2{\newtoks#1\global#1\expandafter{\csname#2\endcsname}}
\def\newread{\alloc@6\read\chardef\sixt@@n}
\def\newwrite{\alloc@7\write\chardef\sixt@@n}
\def\newfam{\alloc@8\fam\chardef\sixt@@n}
\def\newinsert#1{\global\advance\insc@unt by\m@ne
     \ch@ck0\insc@unt\count
     \ch@ck1\insc@unt\dimen
     \ch@ck2\insc@unt\skip
     \ch@ck4\insc@unt\box
     \allocationnumber=\insc@unt
     \global\chardef#1=\allocationnumber
     \wlog{\string#1=\string\insert\the\allocationnumber}}
\def\newif#1{\count@\escapechar \escapechar\m@ne
     \expandafter\expandafter\expandafter
          \xdef\@if#1{true}{\let\noexpand#1=\noexpand\iftrue}%
     \expandafter\expandafter\expandafter
          \xdef\@if#1{false}{\let\noexpand#1=\noexpand\iffalse}%
     \global\@if#1{false}\escapechar=\count@}


\newlinechar=`\^^J
\overfullrule=0pt




\let\itfam=\undefined

\let\bffam=\undefined

\count18=3


\chardef\sharps="19


\mathchardef\alpha="710B
\mathchardef\beta="710C
\mathchardef\gamma="710D
\mathchardef\delta="710E
\mathchardef\epsilon="710F
\mathchardef\zeta="7110
\mathchardef\eta="7111
\mathchardef\theta="7112
\mathchardef\iota="7113
\mathchardef\kappa="7114
\mathchardef\lambda="7115
\mathchardef\mu="7116
\mathchardef\nu="7117
\mathchardef\xi="7118
\mathchardef\pi="7119
\mathchardef\rho="711A
\mathchardef\sigma="711B
\mathchardef\tau="711C
\mathchardef\upsilon="711D
\mathchardef\phi="711E
\mathchardef\chi="711F
\mathchardef\psi="7120
\mathchardef\omega="7121
\mathchardef\varepsilon="7122
\mathchardef\vartheta="7123
\mathchardef\varpi="7124
\mathchardef\varrho="7125
\mathchardef\varsigma="7126
\mathchardef\varphi="7127
\mathchardef\imath="717B
\mathchardef\jmath="717C
\mathchardef\ell="7160
\mathchardef\wp="717D
\mathchardef\partial="7140
\mathchardef\flat="715B
\mathchardef\natural="715C
\mathchardef\sharp="715D



\def\angle{{\vbox{\ialign{$\m@th\scriptstyle##$\crcr
     \not\mathrel{\mkern14mu}\crcr
     \noalign{\nointerlineskip}
     \mkern2.5mu\leaders\hrule height.34\rp@\hfill\mkern2.5mu\crcr}}}}
\def\vdots{\vbox{\baselineskip4\rp@ \lineskiplimit\z@
     \kern6\rp@\hbox{.}\hbox{.}\hbox{.}}}
\def\ddots{\mathinner{\mkern1mu\raise7\rp@\vbox{\kern7\rp@\hbox{.}}\mkern2mu
     \raise4\rp@\hbox{.}\mkern2mu\raise\rp@\hbox{.}\mkern1mu}}
\def\overrightarrow#1{\vbox{\ialign{##\crcr
     \rightarrowfill\crcr
     \noalign{\kern-\rp@\nointerlineskip}
     $\hfil\displaystyle{#1}\hfil$\crcr}}}
\def\overleftarrow#1{\vbox{\ialign{##\crcr
     \leftarrowfill\crcr
     \noalign{\kern-\rp@\nointerlineskip}
     $\hfil\displaystyle{#1}\hfil$\crcr}}}
\def\overbrace#1{\mathop{\vbox{\ialign{##\crcr
     \noalign{\kern3\rp@}
     \downbracefill\crcr
     \noalign{\kern3\rp@\nointerlineskip}
     $\hfil\displaystyle{#1}\hfil$\crcr}}}\limits}
\def\underbrace#1{\mathop{\vtop{\ialign{##\crcr
     $\hfil\displaystyle{#1}\hfil$\crcr
     \noalign{\kern3\rp@\nointerlineskip}
     \upbracefill\crcr
     \noalign{\kern3\rp@}}}}\limits}
\def\big#1{{\hbox{$\left#1\vbox to8.5\rp@ {}\right.\n@space$}}}
\def\Big#1{{\hbox{$\left#1\vbox to11.5\rp@ {}\right.\n@space$}}}
\def\bigg#1{{\hbox{$\left#1\vbox to14.5\rp@ {}\right.\n@space$}}}
\def\Bigg#1{{\hbox{$\left#1\vbox to17.5\rp@ {}\right.\n@space$}}}
\def\@vereq#1#2{\lower.5\rp@\vbox{\baselineskip\z@skip\lineskip-.5\rp@
     \ialign{$\m@th#1\hfil##\hfil$\crcr#2\crcr=\crcr}}}
\def\rlh@#1{\vcenter{\hbox{\ooalign{\raise2\rp@
     \hbox{$#1\rightharpoonup$}\crcr
     $#1\leftharpoondown$}}}}
\def\bordermatrix#1{\begingroup\m@th
     \setbox\z@\vbox{%
          \def\cr{\crcr\noalign{\kern2\rp@\global\let\cr\endline}}%
          \ialign{$##$\hfil\kern2\rp@\kern\p@renwd
               &\thinspace\hfil$##$\hfil&&\quad\hfil$##$\hfil\crcr
               \omit\strut\hfil\crcr
               \noalign{\kern-\baselineskip}%
               #1\crcr\omit\strut\cr}}%
     \setbox\tw@\vbox{\unvcopy\z@\global\setbox\@ne\lastbox}%
     \setbox\tw@\hbox{\unhbox\@ne\unskip\global\setbox\@ne\lastbox}%
     \setbox\tw@\hbox{$\kern\wd\@ne\kern-\p@renwd\left(\kern-\wd\@ne
          \global\setbox\@ne\vbox{\box\@ne\kern2\rp@}%
          \vcenter{\kern-\ht\@ne\unvbox\z@\kern-\baselineskip}%
          \,\right)$}%
     \null\;\vbox{\kern\ht\@ne\box\tw@}\endgroup}
\def\endinsert{\egroup
     \if@mid\dimen@\ht\z@
          \advance\dimen@\dp\z@
          \advance\dimen@12\rp@
          \advance\dimen@\pagetotal
          \ifdim\dimen@>\pagegoal\@midfalse\p@gefalse\fi
     \fi
     \if@mid\bigskip\box\z@
          \bigbreak
     \else\insert\topins{\penalty100 \splittopskip\z@skip
               \splitmaxdepth\maxdimen\floatingpenalty\z@
               \ifp@ge\dimen@\dp\z@
                    \vbox to\vsize{\unvbox\z@\kern-\dimen@}%
               \else\box\z@\nobreak\bigskip
               \fi}%
     \fi
     \endgroup}


\def\cases#1{\left\{\,\vcenter{\m@th
     \ialign{$##\hfil$&\quad##\hfil\crcr#1\crcr}}\right.}
\def\matrix#1{\null\,\vcenter{\m@th
     \ialign{\hfil$##$\hfil&&\quad\hfil$##$\hfil\crcr
          \mathstrut\crcr
          \noalign{\kern-\baselineskip}
          #1\crcr
          \mathstrut\crcr
          \noalign{\kern-\baselineskip}}}\,}


\newif\ifraggedbottom

\def\raggedbottom{\ifraggedbottom\else
     \advance\topskip by\z@ plus60pt \raggedbottomtrue\fi}%
\def\normalbottom{\ifraggedbottom
     \advance\topskip by\z@ plus-60pt \raggedbottomfalse\fi}

\message{hacks,}


\toksdef\toks@i=1
\toksdef\toks@ii=2


\def\TeX{T\kern-.1667em \lower.5ex \hbox{E}\kern-.125em X\null}
\def\jyTeX{{\leavevmode
     \raise.587ex \hbox{\it\j}\kern-.1em \lower.048ex \hbox{\it y}\kern-.12em
     \TeX}}

\let\then=\iftrue
\def\ifnoarg#1\then{\def\hack@{#1}\ifx\hack@\empty}
\def\ifundefined#1\then{%
     \expandafter\ifx\csname\expandafter\blank\string#1\endcsname\relax}
\def\useif#1\then{\csname#1\endcsname}
\def\usename#1{\csname#1\endcsname}
\def\useafter#1#2{\expandafter#1\csname#2\endcsname}

\long\def\loop#1\repeat{\def\@iterate{#1\expandafter\@iterate\fi}\@iterate
     \let\@iterate=\relax}

\let\TeXend=\end
\def\begin#1{\begingroup\def\@@blockname{#1}\usename{begin#1}}
\def\end#1{\usename{end#1}\def\hack@{#1}%
     \ifx\@@blockname\hack@
          \endgroup
     \else\err@badgroup\hack@\@@blockname
     \fi}
\def\@@blockname{}

\def\defaultoption[#1]#2{%
     \def\hack@{\ifx\hack@ii[\toks@={#2}\else\toks@={#2[#1]}\fi\the\toks@}%
     \futurelet\hack@ii\hack@}

\def\markup#1{\let\@@marksf=\empty
     \ifhmode\edef\@@marksf{\spacefactor=\the\spacefactor\relax}\/\fi
     ${}^{\hbox{\subscriptfonts#1}}$\@@marksf}


\newtoks\shortyear
\newtoks\militaryhour
\newtoks\standardhour
\newtoks\minute
\newtoks\amorpm

\def\settime{\count@=\time\divide\count@ by60
     \militaryhour=\expandafter{\number\count@}%
     {\multiply\count@ by-60 \advance\count@ by\time
          \xdef\hack@{\ifnum\count@<10 0\fi\number\count@}}%
     \minute=\expandafter{\hack@}%
     \ifnum\count@<12
          \amorpm={am}
     \else\amorpm={pm}
          \ifnum\count@>12 \advance\count@ by-12 \fi
     \fi
     \standardhour=\expandafter{\number\count@}%
     \def\hack@19##1##2{\shortyear={##1##2}}%
          \expandafter\hack@\the\year}

\def\monthword#1{%
     \ifcase#1
          $\bullet$\err@badcountervalue{monthword}%
          \or January\or February\or March\or April\or May\or June%
          \or July\or August\or September\or October\or November\or December%
     \else$\bullet$\err@badcountervalue{monthword}%
     \fi}

\def\monthabbr#1{%
     \ifcase#1
          $\bullet$\err@badcountervalue{monthabbr}%
          \or Jan\or Feb\or Mar\or Apr\or May\or Jun%
          \or Jul\or Aug\or Sep\or Oct\or Nov\or Dec%
     \else$\bullet$\err@badcountervalue{monthabbr}%
     \fi}

\def\militarytime{\the\militaryhour:\the\minute}
\def\standardtime{\the\standardhour:\the\minute}


\def\@setnumstyle#1#2{\expandafter\global\expandafter\expandafter
     \expandafter\let\expandafter\expandafter
     \csname @\expandafter\blank\string#1style\endcsname
     \csname#2\endcsname}
\def\numstyle#1{\usename{@\expandafter\blank\string#1style}#1}
\def\ifblank#1\then{\useafter\ifx{@\expandafter\blank\string#1}\blank}

\def\blank#1{}

\def\Roman#1{\expandafter\uppercase\expandafter{\romannumeral#1}}
\def\alphabetic#1{%
     \ifcase#1
          $\bullet$\err@badcountervalue{alphabetic}%
          \or a\or b\or c\or d\or e\or f\or g\or h\or i\or j\or k\or l\or m%
          \or n\or o\or p\or q\or r\or s\or t\or u\or v\or w\or x\or y\or z%
     \else$\bullet$\err@badcountervalue{alphabetic}%
     \fi}
\def\Alphabetic#1{\expandafter\uppercase\expandafter{\alphabetic{#1}}}
\def\symbols#1{%
     \ifcase#1
          $\bullet$\err@badcountervalue{symbols}%
          \or*\or\dag\or\ddag\or\S\or$\|$%
          \or**\or\dag\dag\or\ddag\ddag\or\S\S\or$\|\|$%
     \else$\bullet$\err@badcountervalue{symbols}%
     \fi}


\catcode`\^^?=13 \def^^?{\relax}

\def\trimleading#1\to#2{\edef#2{#1}%
     \expandafter\@trimleading\expandafter#2#2^^?^^?}
\def\@trimleading#1#2#3^^?{\ifx#2^^?\def#1{}\else\def#1{#2#3}\fi}

\def\trimtrailing#1\to#2{\edef#2{#1}%
     \expandafter\@trimtrailing\expandafter#2#2^^? ^^?\relax}
\def\@trimtrailing#1#2 ^^?#3{\ifx#3\relax\toks@={}%
     \else\def#1{#2}\toks@={\trimtrailing#1\to#1}\fi
     \the\toks@}

\def\trim#1\to#2{\trimleading#1\to#2\trimtrailing#2\to#2}

\catcode`\^^?=15


\long\def\additemL#1\to#2{\toks@={\^^\{#1}}\toks@ii=\expandafter{#2}%
     \xdef#2{\the\toks@\the\toks@ii}}

\long\def\additemR#1\to#2{\toks@={\^^\{#1}}\toks@ii=\expandafter{#2}%
     \xdef#2{\the\toks@ii\the\toks@}}

\def\getitemL#1\to#2{\expandafter\@getitemL#1\hack@#1#2}
\def\@getitemL\^^\#1#2\hack@#3#4{\def#4{#1}\def#3{#2}}

\message{font macros,}


\newdimen\rp@
\newcount\@@sizeindex \@@sizeindex=0
\newcount\@@factori
\newcount\@@factorii
\newcount\@@factoriii
\newcount\@@factoriv

\countdef\maxfam=18
\newfam\itfam
\newfam\bffam
\newfam\bfsfam
\newfam\bmitfam

\def\@mathfontinit{\count@=4
     \loop\textfont\count@=\nullfont
          \scriptfont\count@=\nullfont
          \scriptscriptfont\count@=\nullfont
          \ifnum\count@<\maxfam\advance\count@ by\@ne
     \repeat}

\def\@fontstyleinit{%
     \def\it{\err@fontnotavailable\it}%
     \def\bf{\err@fontnotavailable\bf}%
     \def\bfs{\err@bfstobf}%
     \def\bmit{\err@fontnotavailable\bmit}%
     \def\sc{\err@fontnotavailable\sc}%
     \def\sl{\err@sltoit}%
     \def\ss{\err@fontnotavailable\ss}%
     \def\tt{\err@fontnotavailable\tt}}

\def\@parameterinit#1{\rm\rp@=.1em \@getscaling{#1}%
     \let\^^\=\@doscaling\scalingskipslist
     \setbox\strutbox=\hbox{\vrule
          height.708\baselineskip depth.292\baselineskip width\z@}}

\def\@getfactor#1#2#3#4{\@@factori=#1 \@@factorii=#2
     \@@factoriii=#3 \@@factoriv=#4}

\def\@getscaling#1{\count@=#1 \advance\count@ by-\@@sizeindex\@@sizeindex=#1
     \ifnum\count@<0
          \let\@mulordiv=\divide
          \let\@divormul=\multiply
          \multiply\count@ by\m@ne
     \else\let\@mulordiv=\multiply
          \let\@divormul=\divide
     \fi
     \edef\@@scratcha{\ifcase\count@                {1}{1}{1}{1}\or
          {1}{7}{23}{3}\or     {2}{5}{3}{1}\or      {9}{89}{13}{1}\or
          {6}{25}{6}{1}\or     {8}{71}{14}{1}\or    {6}{25}{36}{5}\or
          {1}{7}{53}{4}\or     {12}{125}{108}{5}\or {3}{14}{53}{5}\or
          {6}{41}{17}{1}\or    {13}{31}{13}{2}\or   {9}{107}{71}{2}\or
          {11}{139}{124}{3}\or {1}{6}{43}{2}\or     {10}{107}{42}{1}\or
          {1}{5}{43}{2}\or     {5}{69}{65}{1}\or    {11}{97}{91}{2}\fi}%
     \expandafter\@getfactor\@@scratcha}

\def\@doscaling#1{\@mulordiv#1by\@@factori\@divormul#1by\@@factorii
     \@mulordiv#1by\@@factoriii\@divormul#1by\@@factoriv}


\newskip\headskip
\newskip\footskip

\def\typesize=#1pt{\count@=#1 \advance\count@ by-10
     \ifcase\count@
          \@setsizex\or\err@badtypesize\or
          \@setsizexii\or\err@badtypesize\or
          \@setsizexiv
     \else\err@badtypesize
     \fi}

\def\@setsizex{\getixpt
     \def\subsubscriptfonts{\vpt}%
          \def\subsubscriptsize{\vpt\@parameterinit{-8}}%
     \def\subscriptfonts{\viipt}\def\subscriptsize{\viipt\@parameterinit{-4}}%
     \def\footnotefonts{\viiipt}\def\footnotesize{\viiipt\@parameterinit{-2}}%
     \def\smallfonts{\ixpt}\def\smallsize{\ixpt\@parameterinit{-1}}%
     \def\normalfonts{\xpt}\def\normalsize{\xpt\@parameterinit{0}}%
     \def\bigfonts{\xiipt}\def\bigsize{\xiipt\@parameterinit{2}}%
     \def\Bigfonts{\xivpt}\def\Bigsize{\xivpt\@parameterinit{4}}%
     \def\biggfonts{\xviipt}\def\biggsize{\xviipt\@parameterinit{6}}%
     \def\Biggfonts{\xxipt}\def\Biggsize{\xxipt\@parameterinit{8}}%
     \def\tinyfonts{\vpt}\def\tinysize{\vpt\@parameterinit{-8}}%
     \def\HUGEFONTS{\xxvpt}\def\HUGESIZE{\xxvpt\@parameterinit{10}}%
     \normalsize\fixedskipslist}

\def\@setsizexii{\getxipt
     \def\subsubscriptfonts{\vipt}%
          \def\subsubscriptsize{\vipt\@parameterinit{-6}}%
     \def\subscriptfonts{\viiipt}%
          \def\subscriptsize{\viiipt\@parameterinit{-2}}%
     \def\footnotefonts{\xpt}\def\footnotesize{\xpt\@parameterinit{0}}%
     \def\smallfonts{\xipt}\def\smallsize{\xipt\@parameterinit{1}}%
     \def\normalfonts{\xiipt}\def\normalsize{\xiipt\@parameterinit{2}}%
     \def\bigfonts{\xivpt}\def\bigsize{\xivpt\@parameterinit{4}}%
     \def\Bigfonts{\xviipt}\def\Bigsize{\xviipt\@parameterinit{6}}%
     \def\biggfonts{\xxipt}\def\biggsize{\xxipt\@parameterinit{8}}%
     \def\Biggfonts{\xxvpt}\def\Biggsize{\xxvpt\@parameterinit{10}}%
     \def\tinyfonts{\vpt}\def\tinysize{\vpt\@parameterinit{-8}}%
     \def\HUGEFONTS{\xxvpt}\def\HUGESIZE{\xxvpt\@parameterinit{10}}%
     \normalsize\fixedskipslist}

\def\@setsizexiv{\getxiiipt
     \def\subsubscriptfonts{\viipt}%
          \def\subsubscriptsize{\viipt\@parameterinit{-4}}%
     \def\subscriptfonts{\xpt}\def\subscriptsize{\xpt\@parameterinit{0}}%
     \def\footnotefonts{\xiipt}\def\footnotesize{\xiipt\@parameterinit{2}}%
     \def\smallfonts{\xiiipt}\def\smallsize{\xiiipt\@parameterinit{3}}%
     \def\normalfonts{\xivpt}\def\normalsize{\xivpt\@parameterinit{4}}%
     \def\bigfonts{\xviipt}\def\bigsize{\xviipt\@parameterinit{6}}%
     \def\Bigfonts{\xxipt}\def\Bigsize{\xxipt\@parameterinit{8}}%
     \def\biggfonts{\xxvpt}\def\biggsize{\xxvpt\@parameterinit{10}}%
     \def\Biggfonts{\err@sizetoolarge\Biggfonts\HUGEFONTS}%
          \def\Biggsize{\err@sizetoolarge\Biggsize\HUGESIZE}%
     \def\tinyfonts{\vpt}\def\tinysize{\vpt\@parameterinit{-8}}%
     \def\HUGEFONTS{\xxvpt}\def\HUGESIZE{\xxvpt\@parameterinit{10}}%
     \normalsize\fixedskipslist}

\def\subsubscriptfonts{\vpt} \def\subsubscriptsize{\vpt\@parameterinit{-8}}
\def\subscriptfonts{\viipt}  \def\subscriptsize{\viipt\@parameterinit{-4}}
\def\footnotefonts{\viiipt}  \def\footnotesize{\viiipt\@parameterinit{-2}}
\def\smallfonts{\err@sizenotavailable\smallfonts}
                             \def\smallsize{\ixpt\@parameterinit{-1}}
\def\normalfonts{\xpt}       \def\normalsize{\xpt\@parameterinit{0}}
\def\bigfonts{\xiipt}        \def\bigsize{\xiipt\@parameterinit{2}}
\def\Bigfonts{\xivpt}        \def\Bigsize{\xivpt\@parameterinit{4}}
\def\biggfonts{\xviipt}      \def\biggsize{\xviipt\@parameterinit{6}}
\def\Biggfonts{\xxipt}       \def\Biggsize{\xxipt\@parameterinit{8}}
\def\tinyfonts{\vpt}         \def\tinysize{\vpt\@parameterinit{-8}}
\def\HUGEFONTS{\xxvpt}       \def\HUGESIZE{\xxvpt\@parameterinit{10}}

\message{document layout,}


\newtoks\everyoutput \everyoutput={}
\newdimen\depthofpage
\newcount\pagenum \pagenum=0

\newdimen\oddtopmargin  \newdimen\eventopmargin
\newdimen\oddleftmargin \newdimen\evenleftmargin
\newtoks\oddhead        \newtoks\evenhead
\newtoks\oddfoot        \newtoks\evenfoot

\def\topmargin{\afterassignment\@seteventop\oddtopmargin}
\def\leftmargin{\afterassignment\@setevenleft\oddleftmargin}
\def\head{\afterassignment\@setevenhead\oddhead}
\def\foot{\afterassignment\@setevenfoot\oddfoot}

\def\@seteventop{\eventopmargin=\oddtopmargin}
\def\@setevenleft{\evenleftmargin=\oddleftmargin}
\def\@setevenhead{\evenhead=\oddhead}
\def\@setevenfoot{\evenfoot=\oddfoot}

\def\pagenumstyle#1{\@setnumstyle\pagenum{#1}}

\newif\ifdraft
\def\draft{\drafttrue\leftmargin=.5in \overfullrule=5pt }

\newif\ifnumup

\def\outputstyle#1{\global\expandafter\let\expandafter
          \@outputstyle\csname#1output\endcsname
     \usename{#1setup}}

\output={\@outputstyle}

\def\normaloutput{\the\everyoutput
     \global\advance\pagenum by\@ne
     \ifodd\pagenum
          \voffset=\oddtopmargin \hoffset=\oddleftmargin
     \else\voffset=\eventopmargin \hoffset=\evenleftmargin
     \fi
     \advance\voffset by-1in  \advance\hoffset by-1in
     \count0=\pagenum
     \expandafter\shipout\pagebox
     \ifnum\outputpenalty>-\@MM\else\dosupereject\fi}

\newdimen\fullhsize
\newbox\leftpage
\newcount\leftpagenum
\newcount\outputpagenum \outputpagenum=0
\let\leftorright=L

\def\twoupoutput{\the\everyoutput
     \global\advance\pagenum by\@ne
     \if L\leftorright
          \global\setbox\leftpage=\leftline{\pagebox}%
          \global\leftpagenum=\pagenum
          \global\let\leftorright=R%
     \else\global\advance\outputpagenum by\@ne
          \ifodd\outputpagenum
               \voffset=\oddtopmargin \hoffset=\oddleftmargin
          \else\voffset=\eventopmargin \hoffset=\evenleftmargin
          \fi
          \advance\voffset by-1in  \advance\hoffset by-1in
          \count0=\leftpagenum \count1=\pagenum
          \shipout\vbox{\hbox to\fullhsize
               {\box\leftpage\hfil\leftline{\pagebox}}}%
          \global\let\leftorright=L%
     \fi
     \ifnum\outputpenalty>-\@MM
     \else\dosupereject
          \if R\leftorright
               \globaldefs=\@ne\head={\hfil}\foot={\hfil}\globaldefs=\z@
               \null\newpage
          \fi
     \fi}

\def\pagebox{\vbox{\makeheadline\pagebody\makefootline}}

\def\makeheadline{%
     \vbox to\z@{\baselinestretch=\@m
          \vskip\topskip\vskip-.708\baselineskip\vskip-\headskip
          \line{\vbox to\ht\strutbox{}%
               \ifodd\pagenum\the\oddhead\else\the\evenhead\fi}%
          \vss}%
     \nointerlineskip}

\def\pagebody{\vbox to\vsize{%
     \boxmaxdepth\maxdepth
     \ifvoid\topins\else\unvbox\topins\fi
     \depthofpage=\dp255
     \unvbox255
     \ifraggedbottom\kern-\depthofpage\vfil\fi
     \ifvoid\footins
     \else\vskip\skip\footins
          \footnoterule
          \unvbox\footins
          \vskip-\footnoteskip
     \fi}}

\def\makefootline{\baselineskip=\footskip
     \line{\ifodd\pagenum\the\oddfoot\else\the\evenfoot\fi}}


\newskip\abovechapterskip
\newskip\belowchapterskip
\newskip\abovesectionskip
\newskip\belowsectionskip
\newskip\abovesubsectionskip
\newskip\belowsubsectionskip

\def\chapterstyle#1{\global\expandafter\let\expandafter\@chapterstyle
     \csname#1text\endcsname}
\def\sectionstyle#1{\global\expandafter\let\expandafter\@sectionstyle
     \csname#1text\endcsname}
\def\subsectionstyle#1{\global\expandafter\let\expandafter\@subsectionstyle
     \csname#1text\endcsname}

\def\chapter#1{%
     \ifdim\lastskip=17sp \else\chapterbreak\vskip\abovechapterskip\fi
     \@chapterstyle{\ifblank\chapternumstyle\then
          \else\newchapternum=\next\chapternumformat\ \fi#1}%
     \nobreak\vskip\belowchapterskip\vskip17sp }

\def\section#1{%
     \ifdim\lastskip=17sp \else\sectionbreak\vskip\abovesectionskip\fi
     \@sectionstyle{\ifblank\sectionnumstyle\then
          \else\newsectionnum=\next\sectionnumformat\ \fi#1}%
     \nobreak\vskip\belowsectionskip\vskip17sp }

\def\subsection#1{%
     \ifdim\lastskip=17sp \else\subsectionbreak\vskip\abovesubsectionskip\fi
     \@subsectionstyle{\ifblank\subsectionnumstyle\then
          \else\newsubsectionnum=\next\subsectionnumformat\ \fi#1}%
     \nobreak\vskip\belowsubsectionskip\vskip17sp }


\let\TeXunderline=\underline
\let\TeXoverline=\overline
\def\underline#1{\relax\ifmmode\TeXunderline{#1}\else
     $\TeXunderline{\hbox{#1}}$\fi}
\def\overline#1{\relax\ifmmode\TeXoverline{#1}\else
     $\TeXoverline{\hbox{#1}}$\fi}

\def\baselinestretch{\afterassignment\@baselinestretch\count@}
\def\@baselinestretch{\baselineskip=\normalbaselineskip
     \divide\baselineskip by\@m\baselineskip=\count@\baselineskip
     \setbox\strutbox=\hbox{\vrule
          height.708\baselineskip depth.292\baselineskip width\z@}%
     \bigskipamount=\the\baselineskip
          plus.25\baselineskip minus.25\baselineskip
     \medskipamount=.5\baselineskip
          plus.125\baselineskip minus.125\baselineskip
     \smallskipamount=.25\baselineskip
          plus.0625\baselineskip minus.0625\baselineskip}

\def\\{\ifhmode\ifnum\lastpenalty=-\@M\else\hfil\penalty-\@M\fi\fi
     \ignorespaces}
\def\newpage{\vfil\break}

\def\lefttext#1{\par{\@text\leftskip=\z@\rightskip=\centering
     \noindent#1\par}}
\def\righttext#1{\par{\@text\leftskip=\centering\rightskip=\z@
     \noindent#1\par}}
\def\centertext#1{\par{\@text\leftskip=\centering\rightskip=\centering
     \noindent#1\par}}
\def\@text{\parindent=\z@ \parfillskip=\z@ \everypar={}%
     \spaceskip=.3333em \xspaceskip=.5em
     \def\\{\ifhmode\ifnum\lastpenalty=-\@M\else\penalty-\@M\fi\fi
          \ignorespaces}}

\def\beginleft{\par\@text\leftskip=\z@ \rightskip=\centering}
     
\def\beginright{\par\@text\leftskip=\centering\rightskip=\z@ }
     
\def\begincenter{\par\@text\leftskip=\centering\rightskip=\centering}

\def\beginnarrow{\defaultoption[\parindent]\@beginnarrow}
\def\@beginnarrow[#1]{\par\advance\leftskip by#1\advance\rightskip by#1}

\begingroup
\catcode`\[=1 \catcode`\{=11
\gdef\beginignore[\endgroup\bgroup
     \catcode`\e=0 \catcode`\\=12 \catcode`\{=11 \catcode`\f=12 \let\or=\relax
     \let\nd{ignor=\fi \let\}=\egroup
     \iffalse}
\endgroup

\long\def\marginnote#1{\leavevmode
     \edef\@marginsf{\spacefactor=\the\spacefactor\relax}%
     \ifdraft\strut\vadjust{%
          \hbox to\z@{\hskip\hsize\hskip.1in
               \vbox to\z@{\vskip-\dp\strutbox
                    \marginnoteformat
                    \vskip-\ht\strutbox
                    \noindent\strut#1\par
                    \vss}%
               \hss}}%
     \fi
     \@marginsf}


\newtoks\everybye \everybye={\par\vfil}
\outer\def\bye{\the\everybye
     \footnotecheck
     \prelabelcheck
     \streamcheck
     \supereject
     \TeXend}

\message{footnotes,}

\newcount\footnotenum \footnotenum=0
\newskip\footnoteskip
\let\@footnotelist=\empty

\def\footnotenumstyle#1{\@setnumstyle\footnotenum{#1}%
     \useafter\ifx{@footnotenumstyle}\symbols
          \global\let\@footup=\empty
     \else\global\let\@footup=\markup
     \fi}

\def\footnote{\footnotecheck\defaultoption[]\@footnote}
\def\@footnote[#1]{\@footnotemark[#1]\@footnotetext}

\def\footnotemark{\defaultoption[]\@footnotemark}
\def\@footnotemark[#1]{\let\@footsf=\empty
     \ifhmode\edef\@footsf{\spacefactor=\the\spacefactor\relax}\/\fi
     \ifnoarg#1\then
          \global\advance\footnotenum by\@ne
          \@footup{\footnotenumformat}%
          \edef\@@foota{\footnotenum=\the\footnotenum\relax}%
          \expandafter\additemR\expandafter\@footup\expandafter
               {\@@foota\footnotenumformat}\to\@footnotelist
          \global\let\@footnotelist=\@footnotelist
     \else\markup{#1}%
          \additemR\markup{#1}\to\@footnotelist
          \global\let\@footnotelist=\@footnotelist
     \fi
     \@footsf}

\def\footnotetext{%
     \ifx\@footnotelist\empty\err@extrafootnotetext\else\@footnotetext\fi}
\def\@footnotetext{%
     \getitemL\@footnotelist\to\@@foota
     \global\let\@footnotelist=\@footnotelist
     \insert\footins\bgroup
     \footnoteformat
     \splittopskip=\ht\strutbox\splitmaxdepth=\dp\strutbox
     \interlinepenalty=\interfootnotelinepenalty\floatingpenalty=\@MM
     \noindent\llap{\@@foota}\strut
     \bgroup\aftergroup\@footnoteend
     \let\@@scratcha=}
\def\@footnoteend{\strut\par\vskip\footnoteskip\egroup}

\def\footnoterule{\normalfonts
     \kern-.3em \hrule width2in height.04em \kern .26em }

\def\footnotecheck{%
     \ifx\@footnotelist\empty
     \else\err@extrafootnotemark
          \global\let\@footnotelist=\empty
     \fi}

\message{labels,}

\let\@@labeldef=\xdef
\newif\if@labelfile
\newwrite\@labelfile
\let\@prelabellist=\empty

\def\label#1#2{\trim#1\to\@@labarg\edef\@@labtext{#2}%
     \edef\@@labname{lab@\@@labarg}%
     \useafter\ifundefined\@@labname\then\else\@yeslab\fi
     \useafter\@@labeldef\@@labname{#2}%
     \ifstreaming
          \expandafter\toks@\expandafter\expandafter\expandafter
               {\csname\@@labname\endcsname}%
          \immediate\write\streamout{\noexpand\label{\@@labarg}{\the\toks@}}%
     \fi}
\def\@yeslab{%
     \useafter\ifundefined{if\@@labname}\then
          \err@labelredef\@@labarg
     \else\useif{if\@@labname}\then
               \err@labelredef\@@labarg
          \else\global\usename{\@@labname true}%
               \useafter\ifundefined{pre\@@labname}\then
               \else\useafter\ifx{pre\@@labname}\@@labtext
                    \else\err@badlabelmatch\@@labarg
                    \fi
               \fi
               \if@labelfile
               \else\global\@labelfiletrue
                    \immediate\write\sixt@@n{--> Creating file \jobname.lab}%
                    \immediate\openout\@labelfile=\jobname.lab
               \fi
               \immediate\write\@labelfile
                    {\noexpand\prelabel{\@@labarg}{\@@labtext}}%
          \fi
     \fi}

\def\putlab#1{\trim#1\to\@@labarg\edef\@@labname{lab@\@@labarg}%
     \useafter\ifundefined\@@labname\then\@nolab\else\usename\@@labname\fi}
\def\@nolab{%
     \useafter\ifundefined{pre\@@labname}\then
          \undefinedlabelformat
          \err@needlabel\@@labarg
          \useafter\xdef\@@labname{\undefinedlabelformat}%
     \else\usename{pre\@@labname}%
          \useafter\xdef\@@labname{\usename{pre\@@labname}}%
     \fi
     \useafter\newif{if\@@labname}%
     \expandafter\additemR\@@labarg\to\@prelabellist}

\def\prelabel#1{\useafter\gdef{prelab@#1}}

\def\ifundefinedlabel#1\then{%
     \expandafter\ifx\csname lab@#1\endcsname\relax}
\def\useiflab#1\then{\csname iflab@#1\endcsname}

\def\prelabelcheck{{%
     \def\^^\##1{\useiflab{##1}\then\else\err@undefinedlabel{##1}\fi}%
     \@prelabellist}}

\message{equation numbering,}

\newcount\chapternum
\newcount\sectionnum
\newcount\subsectionnum
\newcount\equationnum
\newcount\subequationnum
\newcount\figurenum
\newcount\subfigurenum
\newcount\tablenum
\newcount\subtablenum

\newif\if@subeqncount
\newif\if@subfigcount
\newif\if@subtblcount

\def\newchapternum{\newsectionnum=\z@\@resetnum\chapternum}
\def\newsectionnum{\newsubsectionnum=\z@\@resetnum\sectionnum}
\def\newsubsectionnum{\newequationnum=\z@\newfigurenum=\z@\newtablenum=\z@
     \@resetnum\subsectionnum}
\def\newequationnum{\newsubequationnum=\z@\@resetnum\equationnum}
\def\newsubequationnum{\@resetnum\subequationnum}
\def\newfigurenum{\newsubfigurenum=\z@\@resetnum\figurenum}
\def\newsubfigurenum{\@resetnum\subfigurenum}
\def\newtablenum{\newsubtablenum=\z@\@resetnum\tablenum}
\def\newsubtablenum{\@resetnum\subtablenum}

\def\@resetnum#1{\global\advance#1by1 \edef\next{\the#1\relax}\global#1}

\newchapternum=0

\def\chapternumstyle#1{\@setnumstyle\chapternum{#1}}
\def\sectionnumstyle#1{\@setnumstyle\sectionnum{#1}}
\def\subsectionnumstyle#1{\@setnumstyle\subsectionnum{#1}}
\def\equationnumstyle#1{\@setnumstyle\equationnum{#1}}
\def\subequationnumstyle#1{\@setnumstyle\subequationnum{#1}%
     \ifblank\subequationnumstyle\then\global\@subeqncountfalse\fi
     \ignorespaces}
\def\figurenumstyle#1{\@setnumstyle\figurenum{#1}}
\def\subfigurenumstyle#1{\@setnumstyle\subfigurenum{#1}%
     \ifblank\subfigurenumstyle\then\global\@subfigcountfalse\fi
     \ignorespaces}
\def\tablenumstyle#1{\@setnumstyle\tablenum{#1}}
\def\subtablenumstyle#1{\@setnumstyle\subtablenum{#1}%
     \ifblank\subtablenumstyle\then\global\@subtblcountfalse\fi
     \ignorespaces}

\def\eqnlabel#1{%
     \if@subeqncount
          \newsubequationnum=\next
     \else\newequationnum=\next
          \ifblank\subequationnumstyle\then
          \else\global\@subeqncounttrue
               \newsubequationnum=\@ne
          \fi
     \fi
     \label{#1}{\puteqnformat}(\puteqn{#1})%
     \ifdraft\rlap{\hskip.1in{\tt#1}}\fi}

\let\puteqn=\putlab

\def\equation#1#2{\useafter\gdef{eqn@#1}{#2\eqno\eqnlabel{#1}}}
\def\Equation#1{\useafter\gdef{eqn@#1}}

\def\putequation#1{\useafter\ifundefined{eqn@#1}\then
     \err@undefinedeqn{#1}\else\usename{eqn@#1}\fi}

\def\eqnseriesstyle#1{\gdef\@eqnseriesstyle{#1}}
\def\begineqnseries{\subequationnumstyle{\@eqnseriesstyle}%
     \defaultoption[]\@begineqnseries}
\def\@begineqnseries[#1]{\edef\@@eqnname{#1}}
\def\endeqnseries{\subequationnumstyle{blank}%
     \expandafter\ifnoarg\@@eqnname\then
     \else\label\@@eqnname{\puteqnformat}%
     \fi
     \aftergroup\ignorespaces}

\def\figlabel#1{%
     \if@subfigcount
          \newsubfigurenum=\next
     \else\newfigurenum=\next
          \ifblank\subfigurenumstyle\then
          \else\global\@subfigcounttrue
               \newsubfigurenum=\@ne
          \fi
     \fi
     \label{#1}{\putfigformat}\putfig{#1}%
     {\def\marginnoteformat{\tt}\marginnote{#1}}}

\let\putfig=\putlab

\def\figseriesstyle#1{\gdef\@figseriesstyle{#1}}
\def\beginfigseries{\subfigurenumstyle{\@figseriesstyle}%
     \defaultoption[]\@beginfigseries}
\def\@beginfigseries[#1]{\edef\@@figname{#1}}
\def\endfigseries{\subfigurenumstyle{blank}%
     \expandafter\ifnoarg\@@figname\then
     \else\label\@@figname{\putfigformat}%
     \fi
     \aftergroup\ignorespaces}

\def\tbllabel#1{%
     \if@subtblcount
          \newsubtablenum=\next
     \else\newtablenum=\next
          \ifblank\subtablenumstyle\then
          \else\global\@subtblcounttrue
               \newsubtablenum=\@ne
          \fi
     \fi
     \label{#1}{\puttblformat}\puttbl{#1}%
     {\def\marginnoteformat{\tt}\marginnote{#1}}}

\let\puttbl=\putlab

\def\tblseriesstyle#1{\gdef\@tblseriesstyle{#1}}
\def\begintblseries{\subtablenumstyle{\@tblseriesstyle}%
     \defaultoption[]\@begintblseries}
\def\@begintblseries[#1]{\edef\@@tblname{#1}}
\def\endtblseries{\subtablenumstyle{blank}%
     \expandafter\ifnoarg\@@tblname\then
     \else\label\@@tblname{\puttblformat}%
     \fi
     \aftergroup\ignorespaces}

\message{reference numbering,}

\newcount\referencenum \referencenum=0
\newcount\@@prerefcount \@@prerefcount=0
\newcount\@@thisref
\newcount\@@lastref
\newcount\@@loopref
\newcount\@@refseq
\newdimen\refnumindent
\let\@undefreflist=\empty

\def\referencenumstyle#1{\@setnumstyle\referencenum{#1}}

\def\referencestyle#1{\usename{@ref#1}}

\def\@refsequential{%
     \gdef\@refpredef##1{\global\advance\referencenum by\@ne
          \let\^^\=0\label{##1}{\^^\{\the\referencenum}}%
          \useafter\gdef{ref@\the\referencenum}{{##1}{\undefinedlabelformat}}}%
     \gdef\@reference##1##2{%
          \ifundefinedlabel##1\then
          \else\def\^^\####1{\global\@@thisref=####1\relax}\putlab{##1}%
               \useafter\gdef{ref@\the\@@thisref}{{##1}{##2}}%
          \fi}%
     \gdef\endputreferences{%
          \loop\ifnum\@@loopref<\referencenum
                    \advance\@@loopref by\@ne
                    \expandafter\expandafter\expandafter\@printreference
                         \csname ref@\the\@@loopref\endcsname
          \repeat
          \par}}

\def\@refpreordered{%
     \gdef\@refpredef##1{\global\advance\referencenum by\@ne
          \additemR##1\to\@undefreflist}%
     \gdef\@reference##1##2{%
          \ifundefinedlabel##1\then
          \else\global\advance\@@loopref by\@ne
               {\let\^^\=0\label{##1}{\^^\{\the\@@loopref}}}%
               \@printreference{##1}{##2}%
          \fi}
     \gdef\endputreferences{%
          \def\^^\####1{\useiflab{####1}\then
               \else\reference{####1}{\undefinedlabelformat}\fi}%
          \@undefreflist
          \par}}

\def\beginprereferences{\par
     \def\reference##1##2{\global\advance\referencenum by1\@ne
          \let\^^\=0\label{##1}{\^^\{\the\referencenum}}%
          \useafter\gdef{ref@\the\referencenum}{{##1}{##2}}}}
\def\endprereferences{\global\@@prerefcount=\the\referencenum\par}

\def\beginputreferences{\par
     \refnumindent=\z@\@@loopref=\z@
     \loop\ifnum\@@loopref<\referencenum
               \advance\@@loopref by\@ne
               \setbox\z@=\hbox{\referencenum=\@@loopref
                    \referencenumformat\enskip}%
               \ifdim\wd\z@>\refnumindent\refnumindent=\wd\z@\fi
     \repeat
     \putreferenceformat
     \@@loopref=\z@
     \loop\ifnum\@@loopref<\@@prerefcount
               \advance\@@loopref by\@ne
               \expandafter\expandafter\expandafter\@printreference
                    \csname ref@\the\@@loopref\endcsname
     \repeat
     \let\reference=\@reference}

\def\@printreference#1#2{\ifx#2\undefinedlabelformat\err@undefinedref{#1}\fi
     \noindent\ifdraft\rlap{\hskip\hsize\hskip.1in \tt#1}\fi
     \llap{\referencenum=\@@loopref\referencenumformat\enskip}#2\par}

\def\reference#1#2{{\par\refnumindent=\z@\putreferenceformat\noindent#2\par}}

\def\putref#1{\trim#1\to\@@refarg
     \expandafter\ifnoarg\@@refarg\then
          \toks@={\relax}%
     \else\@@lastref=-\@m\def\@@refsep{}\def\@more{\@nextref}%
          \toks@={\@nextref#1,,}%
     \fi\the\toks@}
\def\@nextref#1,{\trim#1\to\@@refarg
     \expandafter\ifnoarg\@@refarg\then
          \let\@more=\relax
     \else\ifundefinedlabel\@@refarg\then
               \expandafter\@refpredef\expandafter{\@@refarg}%
          \fi
          \def\^^\##1{\global\@@thisref=##1\relax}%
          \global\@@thisref=\m@ne
          \setbox\z@=\hbox{\putlab\@@refarg}%
     \fi
     \advance\@@lastref by\@ne
     \ifnum\@@lastref=\@@thisref\advance\@@refseq by\@ne\else\@@refseq=\@ne\fi
     \ifnum\@@lastref<\z@
     \else\ifnum\@@refseq<\thr@@
               \@@refsep\def\@@refsep{,}%
               \ifnum\@@lastref>\z@
                    \advance\@@lastref by\m@ne
                    {\referencenum=\@@lastref\putrefformat}%
               \else\undefinedlabelformat
               \fi
          \else\def\@@refsep{--}%
          \fi
     \fi
     \@@lastref=\@@thisref
     \@more}

\message{streaming,}

\newif\ifstreaming

\def\streamto{\defaultoption[\jobname]\@streamto}
\def\@streamto[#1]{\global\streamingtrue
     \immediate\write\sixt@@n{--> Streaming to #1.str}%
     \newwrite\streamout\immediate\openout\streamout=#1.str }

\def\streamfrom{\defaultoption[\jobname]\@streamfrom}
\def\@streamfrom[#1]{\newread\streamin\openin\streamin=#1.str
     \ifeof\streamin
          \expandafter\err@nostream\expandafter{#1.str}%
     \else\immediate\write\sixt@@n{--> Streaming from #1.str}%
          \let\@@labeldef=\gdef
          \ifstreaming
               \edef\@elc{\endlinechar=\the\endlinechar}%
               \endlinechar=\m@ne
               \loop\read\streamin to\@@scratcha
                    \ifeof\streamin
                         \streamingfalse
                    \else\toks@=\expandafter{\@@scratcha}%
                         \immediate\write\streamout{\the\toks@}%
                    \fi
                    \ifstreaming
               \repeat
               \@elc
               \input #1.str
               \streamingtrue
          \else\input #1.str
          \fi
          \let\@@labeldef=\xdef
     \fi}

\def\streamcheck{\ifstreaming
     \immediate\write\streamout{\pagenum=\the\pagenum}%
     \immediate\write\streamout{\footnotenum=\the\footnotenum}%
     \immediate\write\streamout{\referencenum=\the\referencenum}%
     \immediate\write\streamout{\chapternum=\the\chapternum}%
     \immediate\write\streamout{\sectionnum=\the\sectionnum}%
     \immediate\write\streamout{\subsectionnum=\the\subsectionnum}%
     \immediate\write\streamout{\equationnum=\the\equationnum}%
     \immediate\write\streamout{\subequationnum=\the\subequationnum}%
     \immediate\write\streamout{\figurenum=\the\figurenum}%
     \immediate\write\streamout{\subfigurenum=\the\subfigurenum}%
     \immediate\write\streamout{\tablenum=\the\tablenum}%
     \immediate\write\streamout{\subtablenum=\the\subtablenum}%
     \immediate\closeout\streamout
     \fi}


\def\err@badtypesize{%
     \errhelp={The limited availability of certain fonts requires^^J%
          that the base type size be 10pt, 12pt, or 14pt.^^J}%
     \errmessage{--> Illegal base type size}}

\def\err@badsizechange{\immediate\write\sixt@@n
     {--> Size change not allowed in math mode, ignored}}

\def\err@sizetoolarge#1{\immediate\write\sixt@@n
     {--> \noexpand#1 too big, substituting HUGE}}

\def\err@sizenotavailable#1{\immediate\write\sixt@@n
     {--> Size not available, \noexpand#1 ignored}}

\def\err@fontnotavailable#1{\immediate\write\sixt@@n
     {--> Font not available, \noexpand#1 ignored}}

\def\err@sltoit{\immediate\write\sixt@@n
     {--> Style \noexpand\sl not available, substituting \noexpand\it}%
     \it}

\def\err@bfstobf{\immediate\write\sixt@@n
     {--> Style \noexpand\bfs not available, substituting \noexpand\bf}%
     \bf}

\def\err@badgroup#1#2{%
     \errhelp={The block you have just tried to close was not the one^^J%
          most recently opened.^^J}%
     \errmessage{--> \noexpand\end{#1} doesn't match \noexpand\begin{#2}}}

\def\err@badcountervalue#1{\immediate\write\sixt@@n
     {--> Counter (#1) out of bounds}}

\def\err@extrafootnotemark{\immediate\write\sixt@@n
     {--> \noexpand\footnotemark command
          has no corresponding \noexpand\footnotetext}}

\def\err@extrafootnotetext{%
     \errhelp{You have given a \noexpand\footnotetext command without first
          specifying^^Ja \noexpand\footnotemark.^^J}%
     \errmessage{--> \noexpand\footnotetext command has no corresponding
          \noexpand\footnotemark}}

\def\err@labelredef#1{\immediate\write\sixt@@n
     {--> Label "#1" redefined}}

\def\err@badlabelmatch#1{\immediate\write\sixt@@n
     {--> Definition of label "#1" doesn't match value in \jobname.lab}}

\def\err@needlabel#1{\immediate\write\sixt@@n
     {--> Label "#1" cited before its definition}}

\def\err@undefinedlabel#1{\immediate\write\sixt@@n
     {--> Label "#1" cited but never defined}}

\def\err@undefinedeqn#1{\immediate\write\sixt@@n
     {--> Equation "#1" not defined}}

\def\err@undefinedref#1{\immediate\write\sixt@@n
     {--> Reference "#1" not defined}}

\def\err@nostream#1{%
     \errhelp={You have tried to input a stream file that doesn't exist.^^J}%
     \errmessage{--> Stream file #1 not found}}

\message{jyTeX initialization}

\everyjob{\immediate\write16{--> jyTeX version \fmtversion}%
     \edef\@@jobname{\jobname}%
     \edef\jobname{\@@jobname}%
     \settime
     \openin0=\jobname.lab
     \ifeof0
     \else\closein0
          \immediate\write16{--> Getting labels from file \jobname.lab}%
          \input\jobname.lab
     \fi}


\def\fixedskipslist{%
     \^^\{\topskip}%
     \^^\{\splittopskip}%
     \^^\{\maxdepth}%
     \^^\{\skip\topins}%
     \^^\{\skip\footins}%
     \^^\{\headskip}%
     \^^\{\footskip}}

\def\scalingskipslist{%
     \^^\{\p@renwd}%
     \^^\{\delimitershortfall}%
     \^^\{\nulldelimiterspace}%
     \^^\{\scriptspace}%
     \^^\{\jot}%
     \^^\{\normalbaselineskip}%
     \^^\{\normallineskip}%
     \^^\{\normallineskiplimit}%
     \^^\{\baselineskip}%
     \^^\{\lineskip}%
     \^^\{\lineskiplimit}%
     \^^\{\bigskipamount}%
     \^^\{\medskipamount}%
     \^^\{\smallskipamount}%
     \^^\{\parskip}%
     \^^\{\parindent}%
     \^^\{\abovedisplayskip}%
     \^^\{\belowdisplayskip}%
     \^^\{\abovedisplayshortskip}%
     \^^\{\belowdisplayshortskip}%
     \^^\{\abovechapterskip}%
     \^^\{\belowchapterskip}%
     \^^\{\abovesectionskip}%
     \^^\{\belowsectionskip}%
     \^^\{\abovesubsectionskip}%
     \^^\{\belowsubsectionskip}}


\def\twoupsetup{
     \topmargin=.75in
     \leftmargin=.5in
     \vsize=6.9in
     \hsize=4.75in
     \fullhsize=10in
     \let\draft=\relax}

\outputstyle{normal}                             

\def\marginnoteformat{\subscriptsize             
     \hsize=1in \baselinestretch=1000 \everypar={}%
     \tolerance=5000 \hbadness=5000 \parskip=0pt \parindent=0pt
     \leftskip=0pt \rightskip=0pt \raggedright}


\head={\ifnumup\normalfonts\hfill\numstyle\pagenum  
        \else\hfil\fi}

\foot={\ifnumup\hfil\else
\hfil\normalfonts\numstyle\pagenum\hfil\fi}  

\normalbaselineskip=12pt                         
\normallineskip=0pt                              
\normallineskiplimit=0pt                         
\normalbaselines                                 

\topskip=.85\baselineskip
\splittopskip=\topskip
\headskip=2\baselineskip
\footskip=\headskip

\pagenumstyle{arabic}                            

\parskip=0pt                                     
\parindent=20pt                                  

\baselinestretch=1000                            


\chapterstyle{left}                              
\chapternumstyle{blank}                          
\def\chapterbreak{\newpage}                      
\abovechapterskip=0pt                            
\belowchapterskip=1.5\baselineskip               
     plus.38\baselineskip minus.38\baselineskip
\def\chapternumformat{\numstyle\chapternum.}     

\sectionstyle{left}                              
\sectionnumstyle{blank}                          
\def\sectionbreak{\vskip0pt plus4\baselineskip\penalty-100
     \vskip0pt plus-4\baselineskip}              
\abovesectionskip=1.5\baselineskip               
     plus.38\baselineskip minus.38\baselineskip
\belowsectionskip=\the\baselineskip              
     plus.25\baselineskip minus.25\baselineskip
\def\sectionnumformat{
     \ifblank\chapternumstyle\then\else\numstyle\chapternum.\fi
     \numstyle\sectionnum.}

\subsectionstyle{left}                           
\subsectionnumstyle{blank}                       
\def\subsectionbreak{\vskip0pt plus4\baselineskip\penalty-100
     \vskip0pt plus-4\baselineskip}              
\abovesubsectionskip=\the\baselineskip           
     plus.25\baselineskip minus.25\baselineskip
\belowsubsectionskip=.75\baselineskip            
     plus.19\baselineskip minus.19\baselineskip
\def\subsectionnumformat{
     \ifblank\chapternumstyle\then\else\numstyle\chapternum.\fi
     \ifblank\sectionnumstyle\then\else\numstyle\sectionnum.\fi
     \numstyle\subsectionnum.}


\footnotenumstyle{symbols}                       
\footnoteskip=0pt                                
\def\footnotenumformat{\numstyle\footnotenum}    
\def\footnoteformat{\footnotesize                
     \everypar={}\parskip=0pt \parfillskip=0pt plus1fil
     \leftskip=1em \rightskip=0pt
     \spaceskip=0pt \xspaceskip=0pt
     \def\\{\ifhmode\ifnum\lastpenalty=-10000
          \else\hfil\penalty-10000 \fi\fi\ignorespaces}}


\def\undefinedlabelformat{$\bullet$}             


\equationnumstyle{arabic}                        
\subequationnumstyle{blank}                      
\figurenumstyle{arabic}                          
\subfigurenumstyle{blank}                        
\tablenumstyle{arabic}                           
\subtablenumstyle{blank}                         

\eqnseriesstyle{alphabetic}                      
\figseriesstyle{alphabetic}                      
\tblseriesstyle{alphabetic}                      

\def\puteqnformat{\hbox{
     \ifblank\chapternumstyle\then\else\numstyle\chapternum.\fi
     \ifblank\sectionnumstyle\then\else\numstyle\sectionnum.\fi
     \ifblank\subsectionnumstyle\then\else\numstyle\subsectionnum.\fi
     \numstyle\equationnum
     \numstyle\subequationnum}}
\def\putfigformat{\hbox{
     \ifblank\chapternumstyle\then\else\numstyle\chapternum.\fi
     \ifblank\sectionnumstyle\then\else\numstyle\sectionnum.\fi
     \ifblank\subsectionnumstyle\then\else\numstyle\subsectionnum.\fi
     \numstyle\figurenum
     \numstyle\subfigurenum}}
\def\puttblformat{\hbox{
     \ifblank\chapternumstyle\then\else\numstyle\chapternum.\fi
     \ifblank\sectionnumstyle\then\else\numstyle\sectionnum.\fi
     \ifblank\subsectionnumstyle\then\else\numstyle\subsectionnum.\fi
     \numstyle\tablenum
     \numstyle\subtablenum}}


\referencestyle{sequential}                      
\referencenumstyle{arabic}                       
\def\putrefformat{\numstyle\referencenum}        
\def\referencenumformat{\numstyle\referencenum.} 
\def\putreferenceformat{
     \everypar={\hangindent=1em \hangafter=1 }%
     \def\\{\hfil\break\null\hskip-1em \ignorespaces}%
     \leftskip=\refnumindent\parindent=0pt \interlinepenalty=1000 }


\normalsize


\def\fmtversion{2.6M (June 1992)}
%
%
%
\def\JSP#1#2#3{{\sl J. Stat. Phys.} {\bf #1} (#2) #3}
\def\PRL#1#2#3{{\sl Phys. Rev. Lett.} {\bf#1} (#2) #3}
\def\PR#1#2#3{{\sl Phys. Rev.} {\bf#1} (#2) #3}

\def\NPB#1#2#3{{\sl Nucl. Phys.} {\bf B#1} (#2) #3}

\def\PRB#1#2#3{{\sl Phys. Rev.} {\bf B#1} (#2) #3}

\def\PLA#1#2#3{{\sl Phys. Lett.} {\bf #1A} (#2) #3}
\def\JMP#1#2#3{{\sl J. Math. Phys.} {\bf #1} (#2) #3}

\def\PTP#1#2#3{{\sl Prog. Theor. Phys.} {\bf #1} (#2) #3}

\def\TMP#1#2#3{{\sl Theor. Mat. Phys.} {\bf #1} (#2) #3}
\def\JPA#1#2#3{{\sl J. Physics} {\bf A#1} (#2) #3}

\def\JSM#1#2#3{{\sl J. Soviet Math.} {\bf #1} (#2) #3}

\def\ZP#1#2#3{{\sl Z.Phys.} {\bf #1} (#2) #3}

\def\upref#1/{\markup{[\putref{#1}]}}
\def\Idoubled#1{{\rm I\kern-.22em #1}}
\def\Odoubled#1{{\setbox0=\hbox{\rm#1}%
     \dimen@=\ht0 \dimen@ii=.04em \advance\dimen@ by-\dimen@ii
     \rlap{\kern.26em \vrule height\dimen@ depth-\dimen@ii width.075em}\box0}}
\def\Real{\Idoubled R}

\catcode`\@=12

\typesize=12pt
\def\half{{1\over 2}}

\def\half{{1\over 2}}
\def\sgn{{\rm sgn}}
\def\upref#1/{\markup{[\putref{#1}]}}
\def\Idoubled#1{{\rm I\kern-.22em #1}}
\def\Odoubled#1{{\setbox0=\hbox{\rm#1}%
     \dimen@=\ht0 \dimen@ii=.04em \advance\dimen@ by-\dimen@ii
     \rlap{\kern.26em \vrule height\dimen@ depth-\dimen@ii width.075em}\box0}}
\def\Real{\Idoubled R}

\def\da{\downarrow}
\catcode`\@=12
\line{\smallfonts\it December 1994\hfil BONN-TH-94-13}
\line{\smallfonts\it \hfil cond-mat/9406081}
\vskip2em
\baselineskip=24pt
\begin{center}
{\bigsize{\sc A Note on Dressed S-Matrices\\ in Models with long-range
    Interactions}}
\end{center}
\baselineskip=16pt
\vskip 1.5cm
\begin{center}
{\bigsize
Fabian H.L.E\sharps ler\footnote[$\ \flat$]{e-mail:
fabman@avzw03.physik.uni-bonn.de}}\vskip 0.3cm
{\it Physikalisches Institut der Universit\"at Bonn\vskip 3pt
Nussallee 12, 53115 Bonn, Germany}

\end{center}

\vskip 1.5cm
\centertext{\bfs \bigsize ABSTRACT}
\vskip\belowsectionskip
\begin{narrow}[4em]
\noindent
\baselineskip=20pt
The {\sl dressed} Scattering matrix describing scattering of quasiparticles
in various models with long-range interactions is evaluated by means
of Korepin's method\upref vek1/.
For models with ${1\over\sin^2(r)}$-interactions the S-matrix is found to
be a momentum-independent phase, which clearly demonstrates the ideal gas
character of the quasiparticles in such models.
We then determine S-matrices for some models with
${1\over\sinh^2(r)}$-interaction and find them to be in general
nontrivial. For the ${1\over r^2}$-limit of the
${1\over\sinh^2(r)}$-interaction we recover trivial S-matrices, thus
exhibiting a crossover from interacting to noninteracting quasiparticles.
The relation of the S-matrix to fractional statistics is discussed.
\end{narrow}
\vskip .5cm
{\sc Pacs: 75.10.Jm , 05.50.q}
\vfil

\break


%
%
%
\def\JSP#1#2#3{{\sl J. Stat. Phys.} {\bf #1} (#2) #3}
\def\PRL#1#2#3{{\sl Phys. Rev. Lett.} {\bf#1} (#2) #3}
\def\PR#1#2#3{{\sl Phys. Rev.} {\bf#1} (#2) #3}

\def\NPB#1#2#3{{\sl Nucl. Phys.} {\bf B#1} (#2) #3}

\def\PRB#1#2#3{{\sl Phys. Rev.} {\bf B#1} (#2) #3}

\def\PLA#1#2#3{{\sl Phys. Lett.} {\bf #1A} (#2) #3}
\def\JMP#1#2#3{{\sl J. Math. Phys.} {\bf #1} (#2) #3}

\def\PTP#1#2#3{{\sl Prog. Theor. Phys.} {\bf #1} (#2) #3}

\def\TMP#1#2#3{{\sl Theor. Mat. Phys.} {\bf #1} (#2) #3}
\def\JPA#1#2#3{{\sl J. Physics} {\bf A#1} (#2) #3}

\def\JSM#1#2#3{{\sl J. Soviet Math.} {\bf #1} (#2) #3}

\def\ZP#1#2#3{{\sl Z.Phys.} {\bf #1} (#2) #3}

%
%

\equation{Hcs}{H= -\sum_{j=1}^N {\partial^2\over\partial
x_j^2}+\sum_{j<k} {2\lambda (\lambda -1)\over (x_k-x_j)^2}\ .}

\equation{Hcsh}{H= -\sum_{j=1}^N {\partial^2\over\partial
x_j^2}+\sum_{j<k} {2\lambda (\lambda -1)\over\sinh^2({x_k-x_j\over
a})}\ .}

\equation{Hcshx}{H= -\sum_{j=1}^N {\partial^2\over\partial
x_j^2}+\sum_{j<k}^N \kappa^2{(\lambda^2 -\lambda
P_{jk})\over\sinh^2({(x_k-x_j)\kappa})}\ .}

\equation{Hlatt}{H_{latt}= -{1\over
2}\sum_{j>k}{1+P_{jk}\over\sinh^2({k-j\over d})}\ .}

\equation{baecs}{\exp(-ik_jL) = \prod_{l\neq j} S(k_j, k_l)\ ,}

\equation{cslog}{L k_j = 2\pi I_j + (\lambda -1) \pi\sum_{l\neq j}
{\rm sgn}(k_j-k_l)\ , \quad j+1\ldots N\ \ ,}

\equation{cshlog}{L k_j = 2\pi I_j + \sum_{l\neq j} \theta(k_j-k_l)\ ,
\quad j+1\ldots N\ \ ,}

\equation{cshlogx}{\eqalign{L k_j &= 2\pi I_j + \sum_{l\neq j} \theta(k_j-k_l)
-\sum_s \vartheta({k_j-\alpha^n_s\over n\lambda})\ , \cr
0 &= 2\pi J^n_s - \sum_{l} \vartheta({\alpha_s-k_l\over n\lambda})
+\sum_{(m,t)\neq
(n,s)}\vartheta_{nm}({\alpha^n_s-\alpha^m_t\over\lambda})\ ,\cr}}

\equation{baespin}{\eqalign{L \xi^{(0)}_j&=\sum_l \theta_0(\xi_j^{(0)}
- \xi_l^{(0)})\cr
0&=2\pi I_j-\sum_l \theta^\prime_0 (\xi_j^{(0)}-\xi_l^{(0)})\xi_l^{(1)}
-\sum_{(n,s)}{\vartheta}\left({2\over n}(\xi_j^{(0)}-\beta^n_s)\right)
+\sum_l\theta_1(\xi_j^{(0)}-\xi_l^{(0)})\cr
0&=2\pi J^n_s-
\sum_l {\vartheta}\left({2\over n}(\beta^n_s-\xi_l^{(0)})\right)
+ \sum_{(m,t)\neq (n,s)}\vartheta_{nm}(2(\beta^n_s-\beta^m_t))\ ,\cr}}

\equation{baespinT}{\eqalign{
0&=2\pi I_j-\sum_l \theta^\prime_0 (\xi_j^{(0)} -
\xi_l^{(0)}){\tilde \xi}_l^{(1)}
-\sum_{s}{\vartheta}\left(2(\xi_j^{(0)}-{\tilde\beta}_s)\right)
+\sum_l\theta_1(\xi_j^{(0)}-\xi_l^{(0)})\cr
&\qquad +\sum_{j=1}^2 {\vartheta}(2(\xi_j^{(0)}-\beta_{h,j})) -\pi\cr
0&=2\pi {\tilde J}_s- \sum_l {\vartheta}(2({\tilde\beta}_s-\xi_l^{(0)}))
+ \sum_t{\vartheta}({\tilde\beta}_s-{\tilde\beta_t})+\pi
-\sum_{j=1}^2 {\vartheta}({\tilde\beta}_s-\beta_{h,j}))\ .\cr}}

\equation{ftsinh}{\eqalign{F_2(\beta)&=1-{1\over 2\pi}\int_{-\infty}^\infty
d\beta^\prime {2\over 1+(\beta-\beta^\prime)^2} F_2(\beta^\prime)
-{1\over 2\pi}\sum_{j=1}^2{\vartheta}(\beta-\beta_{h,j})\cr
0&=-\int_{-a}^a d\xi^\prime\ F_1(\xi^\prime)\theta_0^\prime(\xi-\xi^\prime)
+\int_{-\infty}^\infty d\beta {2\over 1+4(\beta-\xi)^2} F_2(\beta)
-\pi+\sum_{j=1}^2{\vartheta}(2(\xi-\beta_{h,j}))\ .\cr}}

\equation{srsgs}{\eqalign{{1\over d}&= \int_{-a}^a d\xi^\prime\
\rho(\xi^\prime)\theta_0^\prime(\xi-\xi^\prime)\ ,\cr
0&= 2\pi\rho(\xi) - \int_{-a}^a\!\! d\xi^\prime\ \gamma(\xi^\prime)
\theta_0^{\prime\prime}(\xi-\xi^\prime) + \int_{-a}^a\!\! d\xi^\prime\
\rho(\xi^\prime) \theta_1^{\prime}(\xi-\xi^\prime)
-\int_{-\infty}^\infty\!\! d\beta {2\sigma(\beta)\over 1+4(\xi-\beta)^2}\cr
0&= 2\pi\sigma(\beta) - \int_{-a}^a d\xi\ \rho(\xi) {2\over 1+4(\xi-\beta)^2}
+-\int_{-\infty}^\infty d\beta^\prime\ \sigma(\beta^\prime)
{2\over 1+(\beta-\beta^\prime)^2}\ ,\cr}}

\equation{cslogx}{\eqalign{L k_j &= 2\pi I_j + \sum_{l\neq j}
\pi(\lambda-1)\sgn(k_j-k_l) -\sum_s \pi\sgn({k_j-\alpha_s})\ , \cr
0 &= 2\pi J_s - \sum_{l} \pi\sgn({\alpha_s-k_l})+\sum_{t\neq
s}\pi\sgn({\alpha_s-\alpha_t\over 2})\ .\cr}}

\equation{Shs}{S_{HS}(k^h_1, k^h_2) = i\ {\rm id}\ .}

\equation{Fs}{F_S(k) = -{\pi\over 2}\sum_{j=1}^2 {\rm sgn}(k-k^h_j) +
\pi\ {\rm sgn}(k- {\kappa})\ .}

\equation{dt}{\delta_T(k^h_1, k^h_2) = 2\pi
F_T(k^h_1)\biggr|_{k^h_1>k^h_2}  = \pi -{\pi\over 2} {\rm
sgn}(k^h_1-k^h_2) = {\pi\over 2}\ .}

\equation{Ft}{F_T(k) = \half - {1\over 4}\sum_{j=1}^2 {\rm
sgn}(k-k^h_j)\ .}

\equation{ft}{F_T(k) = 1- \int_{-\pi}^\pi dk^\prime \delta(k-k^\prime)
F_T(k^\prime) -{1\over 2} \sum_{j=1}^2 {\rm sgn}(k-k^h_j)\ .}

\equation{baest}{N {\tilde k}^1_\alpha = 2\pi {\tilde I}^1_\alpha +
\pi\sum_{\scriptstyle\beta=1\atop\scriptstyle\beta\neq\alpha}^{{N\over
2}-1}  {\rm sgn}({\tilde k}^1_\alpha -{\tilde k}^1_\beta)
-\pi\sum_{j=1}^2 {\rm sgn}({\tilde k}^1_\alpha - k^h_j)\ .}

\equation{rho}{{1\over 2\pi} = \rho_1(k) + \int_{-\pi}^\pi dk^\prime
\delta(k-k^\prime) \rho_1(k^\prime)\ ,}

\equation{baegs}{ N k^1_\alpha = 2\pi I^1_\alpha
+\pi\sum_{\scriptstyle\beta =1\atop\scriptstyle\beta\neq\alpha}^{M} {\rm
sgn}(k^1_\alpha -k^1_\beta)\ ,\quad \alpha =1\ldots M.}

\equation{int}{|I^n_\alpha|\leq {1\over 2} (N-\sum_{m=1}^\infty t_{nm}
M_m -1)\ ,}

\equation{intx}{|J^n_s|\leq {1\over 2} (N-\sum_{m=1}^\infty t_{nm}
M_m -1)\ ,}

\equation{baehs}{Nk^n_\alpha = 2\pi I^n_\alpha + \pi\sum_{m\beta}
t_{nm}  {\rm sgn}(k^n_\alpha - k^m_\beta)\ ,}

\equation{Hhs}{H = 2\sum_{i<j} {1\over( {N\over\pi}\sin({\pi\over
N}(i-j)))^2}  (P_{ij}-1)\ ,}

\equation{su3}{\eqalign{&Nk^{(1)n}_\alpha = 2\pi I^{(1)n}_\alpha +
\pi\sum_{m\beta} t_{nm} {\rm \ sgn}(k^{(1)n}_\alpha - k^{(1)m}_\beta)
- \pi\sum_{m\beta} {\rm min}(n,m) {\rm \ sgn}(k^{(1)n}_\alpha -
k^{(2)m}_\beta)\cr
&\pi\sum_{m\beta} {\rm min}(n,m) {\rm \ sgn}(k^{(2)n}_\alpha -
k^{(1)m}_\beta)= 2\pi I^{(2)n}_\alpha +
\pi\sum_{m\beta} t_{nm} {\rm \ sgn}(k^{(2)n}_\alpha - k^{(2)m}_\beta)\
.\cr}}

\equation{dens}{\rho^{(1)}_1(k) = {1\over 3\pi},\quad ,\
\rho^{(2)}_1(k) = {1\over 6\pi}  \ .}

\equation{oct}{\eqalign{N{\tilde k}^{(1)1}_\alpha &= 2\pi {\tilde
I}^{(1)1}_\alpha + \pi\sum_{\beta=1}^{{2N\over3}} {\rm \
sgn}({\tilde k}^{(1)1}_\alpha - {\tilde k}^{(1)1}_\beta) -
\pi\sum_{\beta=1}^{{N\over 3}} {\rm \ sgn}({\tilde k}^{(1)1}_\alpha -
{\tilde k}^{(2)1}_\beta)\cr
&\hskip 100pt - \pi{\rm \ sgn}({\tilde k}^{(1)1}_\alpha -
k^{(1)}_h)  +\pi{\rm \ sgn}({\tilde k}^{(1)1}_\alpha - k^{(2)}_h) \cr
\pi\sum_{\beta=1}^{{2N\over 3}} {\rm \ sgn}({\tilde k}^{(2)1}_\alpha -
{\tilde k}^{(1)1}_\beta)&= 2\pi {\tilde I}^{(2)1}_\alpha +
\pi\sum_{\beta=1}^{{N\over 3}} {\rm \ sgn}({\tilde k}^{(2)1}_\alpha -
{\tilde k}^{(2)1}_\beta)\cr
&\hskip 100pt + \pi{\rm \ sgn}({\tilde k}^{(2)1}_\alpha -
k^{(1)}_h)  -\pi{\rm \ sgn}({\tilde k}^{(2)1}_\alpha - k^{(2)}_h)\
.\cr}}

\equation{cshkern}{\theta(k) = i\ \ln\left({\Gamma(1+{ik\over 2})\over
\Gamma(1-{ik\over 2})}{\Gamma(\lambda-{ik\over 2})\over
\Gamma(\lambda+{ik\over 2})}\right)\ .}

\equation{cshshift}{\delta(k_1,k_2)+{1\over 2\pi}\int_{-B}^B dk\
\theta^\prime(k_1-k)\ \delta(k,k_2) = \pi + \theta(k_1-k_2)\ .}

\equation{Ftx}{\eqalign{F^T_1(k)&= {1\over 2\pi}\sum_{p=1}^2\left(
2\ \arctan\left(e^{\pi{k-\alpha_{h,p}\over 2\lambda}}\right)-{\pi\over
2}\right) - {1\over 2\pi}\int_{-A}^Adk^\prime\
F^T_1(k^\prime)\Theta(k-k^\prime)\ ,\cr
F^T_2(\alpha)&= {1\over 2}-{i\over 2\pi}\sum_{p=1}^2 \ln\left({
\Gamma\left({1+i{\alpha-\alpha_{h,p}\over 2\lambda}\over 2}\right)
\Gamma\left(1-i{\alpha-\alpha_{h,p}\over 4\lambda}\right)\over
\Gamma\left({1-i{\alpha-\alpha_{h,p}\over 2\lambda}\over 2}\right)
\Gamma\left(1+i{\alpha-\alpha_{h,p}\over 4\lambda}\right)}\right) +
{1\over 4\lambda}\int_{-A}^A\!\!dk {F^T_1(k)\over \cosh\left({\pi\over
2\lambda} (\alpha-k)\right)}\ ,\cr}}

\equation{Fsx}{\eqalign{F^S_1(k) &= F^T_1(k)\ ,\cr
F^S_2(\alpha)&= F^T_2(\alpha)+{1\over \pi}\arctan\left({\alpha -
{\alpha_{h,1}+\alpha_{h,2}\over 2}\over\lambda}\right) - \half\ ,\cr}}

\equation{smx}{S(\alpha_{h,1}, \alpha_{h,2}) =
e^{2\pi iF^T_2(\alpha_{h,1})}\left({\nu\over\nu+i}{\rm id} +
{i\over\nu+i}P\right)\ ,\nu={\alpha_{h,1}-\alpha_{h,2}\over2\lambda}>0,}

\equation{sxxx}{S_{XXX}(\mu) = -{\Gamma({1+i\mu\over
2})\Gamma(1-{i\mu\over 2})\over \Gamma({1-i\mu\over
2})\Gamma(1+{i\mu\over 2})}\left({\mu\over \mu+i}{\rm id}\ +{i\over\mu
+i} P\right)\ ,}

\equation{cshaba}{\eqalign{e^{ik_jL}&= \prod_{l\neq j}^N
{\Gamma(1-{i(k_j-k_l)\over 2})\over \Gamma(1+{i(k_j-k_l)\over
2})}{\Gamma(\lambda+{i(k_j-k_l)\over 2})\over
\Gamma(\lambda-{i(k_j-k_l)\over 2})}
\prod_{s=1}^{N_\da}{k_j-\alpha_s+i\lambda\over k_j-\alpha_s-i\lambda}\
,\cr
\prod_{j=1}^N{\alpha_s-k_j+i\lambda\over \alpha_s-k_j-i\lambda}&=
-\prod_{j=1}^{N_\da}{\alpha_s-\alpha_t+2i\lambda\over
\alpha_s-\alpha_t-2i\lambda}\ .\cr}}

%
%
\pagenumstyle{blank}
\sectionnumstyle{arabic}
\footnoteskip=2pt

\vskip 1cm
\pagenumstyle{arabic}
\baselineskip=16pt
\sectionnum=0

{\sc\section{Introduction}} Haldane\upref h1/ recently has put forward
an interpretation of the Haldane-Shastry model as a generalized ideal
gas with fractional statistics (``ideal gas of spinons''). The
Calogero-Sutherland (CS) model is another example for a system of free
particles with fractional statistics\upref p1, bw, pas, ha, p2/.
All methods employed so far in exhibiting the ideal gas character and
the nature of the statistics are based on the knowledge of the exact
wave-functions\upref bghp/ for the HS and CS models. For the multitude
of other models with long-range iteractions, in particular models
solvable by Asymptotic Bethe Ansatz (ABA)\upref s1,s2/, exact
wave-functions are not known. It would be useful to have a method
based merely on the ABA to decide whether or not those systems fall
into Haldane's category of ideal gases with fractional statistics. The
most direct way to determine whether a system of quasiparticles is an
ideal gas is to evaluate the dressed scattering matrix describing
scattering of the elementary excitations in the model. If it is a
momentum-independent phase, then we are indeed dealing with an ideal
gas. Furthermore, if the phase it not $\pm 1$, the quasiparticles have
fractional statistics in the sense that the phase of the wave-function
under interchange of two particles is neither bosonic ($+1$) nor
fermionic ($-1$). This follows from the observation that for
noninteracting particles ({\sl i.e.} for momentum-independent
S-matrices) the scattering phase is precisely equal to the phase
picked up by interchanging the two particles.

The plan of this note is as follows: in sections $2$ and $3$ we
determine the S-matrix for scattering of quasiparticles in $SU(N)$
Haldane-Shastry chains. It is found to be a momentum-independent
phase, which shows both the ideal-gas character and the fractional
statistics of the quasiparticles. In section $4$ we repeat this
analysis for the case of the Calogero-Sutherland model. In
section $5$ we discuss the generalization of our results to other
models with ${1\over\sin^2(r)}$-interactions. In section $6$ we
consider the ${1\over\sinh^2(r)}$-CS model and its exchange
generalizations and show that quasiparticles in these models are {\sl
interacting}, and become free in the ${1\over r^2}$ limit
only. In section $7$ we summarize and discuss our results.

{\sc\section{$SU(2)$ Haldane-Shastry Chain}}
\noindent
The hamiltonian of the $SU(2)$ HS chain is given by\upref h2,sh/
$$\putequation{Hhs}$$
where $P_{ij}$ is a permutation operator exchanging the spins at sites
$i$ and $j$. Ha and Haldane\upref hh/ proposed to characterize
eigenstates of (\putlab{Hhs}) by means of sets of spectral variables
$k^n_\alpha$ obeying the following set of ``Bethe-like equations''
$$\putequation{baehs}$$
where $k^n_\alpha$ is the position of the center of a ``string of length n''
($\alpha$ labels different strings of the same length), $t_{nm} =
2{\rm min}(n,m)-\delta_{mn}$, and $I^n_\alpha$ are integer or half-odd integer
quantum numbers with range
$$\putequation{int}$$
where $M_m$ is the number of strings of given length $m$. Thus
$\sum_{n=1}^\infty n M_n =M$, where $M$ is the total number of down spins.
Energy and momentum are given as $E=\sum_{n\alpha} \half\left[
(k^{n}_\alpha )^2 - \pi^2\right]$ and $P=\sum_{n\alpha}
(k^{n}_\alpha + \pi) $.
The above equations are very similar to the Bethe equations for the
spin-$1\over 2$ Heisenberg XXX (nearest neighbour)
antiferromagnet\upref HaBe, ft1,ft2,mt/. Ha and Haldane proceed to
show that ground state and excitations as well as the thermodynamics
of the HS chain are correctly described by the above equations if one
considers them as ``true Bethe equations'' for an integrable system.
The ground state is a filled ``Fermi-sea'', where all
vacancies for the integers $I^1_\alpha$ (allowed by (\putlab{int}))
are taken. More precisely we have $M=M_1={N\over 2}$, and there are
$N\over 2$ vacancies for the integers $I^1_\alpha$, all of which are
filled. This corresponds to filling all vacancies for the momenta
$k^1_\alpha$ between $-\pi$ and $\pi$. The Ha-Haldane equations take
the form
$$\putequation{baegs}$$
Subtracting (\putlab{baegs}) for $\alpha$ and $\alpha +1$ one obtains
an equation for the density of $k$'s $\rho_1(k^1_\alpha) = {1\over
  k^1_{\alpha +1} - k^1_{\alpha}}$, which in the thermodynamic limit
$N\rightarrow\infty$ turns into the following integral equation
$$\putequation{rho}$$
which can be solved trivially with the result $\rho_1(k)={1\over 4\pi}$.
This shows that the ground state is of much simpler nature than for the
Heisenberg antiferromagnet.

The elementary excitations or ``quasiparticles'' are
identified as two spin-$1\over 2$ objects, called spinons, with
dispersion (using the conventions of [\putref{hh}])
$\epsilon(p) = p(\pi -p),\ p\in [0,\pi]$. The situation is thus very
similar to the nearest neighbour Heisenberg model, where there are
also two elementary excitations\upref ft1, ft2/ carrying spin $1\over
2$, but with dispersion\footnote{In order to compare the results of
[\putref{ft1}] with the Haldane-Shastry case we should set $J=2$ in
the hamiltonian of [\putref{ft1}].} $\epsilon_{_{XXX}}(p) =
\pi\sin(p),\ p\in [0,\pi]$. The similarity is not surprising due to
the fact that (\putlab{int}) are the same for both models, and the
ground states of both models are given by filling a Fermi sea of real
spectral parameters.
The $SU(2)$ structure of excitations is the same for the HS
chain\upref h4/ and the nearest neighbour model: all excited states
over the true ground state are {\sl scattering states} of an {\sl even
number} of quasiparticles. A scattering state is characterized by
having energy and momentum equal to the sum of the energies/momenta of
its constituent quasiparticles. The simplest excited states are in the
two-particle sector, and their $SU(2)$-representation theory is simply
given by tensoring two fundamental representations (the quasiparticles
transform in the fundamental representation $\half$):
$\half\otimes\half = 1\oplus 0$. In other words there are four excited
states in the two-particle sector, three of which form a $SU(2)$
triplet and one a $SU(2)$ singlet. Their energies/momenta are
degenerate and given by the sum of the quasiparticle energies/momenta
$E=\epsilon(p_1)+\epsilon(p_2)$, $P=p_1+p_2$. In the four-particle
sector we obtain the $SU(2)$ representation content
$\half\otimes\half\otimes\half\otimes\half = 2\oplus 1\oplus 1\oplus
1\oplus 0\oplus 0$ and so on.

The dressed S-matrix can also be obtained from the equations
(\putlab{baehs}) and (\putlab{int}), if we treat the Ha-Haldane
equations the same way we treat Bethe equations for Bethe-Ansatz
solvable models. This is not as straightforward as it may seem, as
Bethe Ansatz equations have a {\sl direct} connection to the {\sl
exact} eigenfunctions of the hamiltonian, which is something missing
for the case of the HS chain and the Ha-Haldane equations. However,
this connection is not vital for determining the scattering
phase-shifts\upref vek1,al/. Adopting (\putlab{baehs}) as ``Bethe
equations'' Korepin's method\upref vek1/ can then be applied in a way
completely analogous to the nearest neighbour XXX chain\upref
ft1,ft2/, for a more detailed explanation of the method we use we
refer to [\putref{vek1, vek2}].

In order to determine the two-quasiparticle scattering matrix we need
to determine the two-quasiparticle eigenstates of the hamiltonian (the
triplet and the singlet), which by construction are also eigenstates
of the scattering operator we seek. We proceed again in a way
analogous to the XXX case: the spin-triplet $SU(2)$-highest weight
state is obtained by choosing $M_1={N\over 2}-1$. The allowed range of
integers is $|I^1_\alpha|\leq {N\over 4}$, which means that there are
${N\over 2}+1$ vacancies and thus $2$ holes. We take the holes to have
momenta $k^h_1$ and $k^h_2$. The Ha-Haldane equations for this
excitation are
$$\putequation{baest}$$
Here we have used the notation $\tilde k$ to indicate that the momenta
of the particles in the Fermi sea are slightly different for the excited
state as compared to the ground state\upref vek2/. Subtracting
(\putlab{baest}) from the equation (\putlab{baegs}) for the ground
state and taking the thermodynamic limit we obtain an integral
equation for the {\sl shift function} $F_T(k)$\upref vek3/ of the
spin-triplet state (which is the limit of the finite-lattice quantity
$F_T(k^1_\alpha) =  {{\tilde k}^1_\alpha - k^1_\alpha\over k^1_{\alpha
    +1}-k^1_\alpha}$)
$$\putequation{ft}$$
The solution to this equation is
$$\putequation{Ft}$$
The phase shift for the spin-triplet state is (for an explanation see
{\sl e.g.} [\putref{vek2}])
$$\putequation{dt}$$
Here we have used that $k^h_1$ must be larger than $k^h_2$ (or vice
versa, in which case $\delta_T(k^h_1, k^h_2) = 2\pi
F_T(k^h_2)\biggr|_{k^h_2>k^h_1}$) for scattering to occur.
The spin-singlet excitation is constructed by taking $M_1 = {N\over 2}-2$,
$M_2 =1$. Now there are $N\over 2$ vacancies for the 1-strings and thus
again $2$ holes, whereas there is only $1$ vacancy for the 2-string.
Denoting the positions of the $2$ holes again by $k^h_j$ and the position
of the 2-string by $\kappa$, we find that the Bethe-equation for the 2-string
leads to the condition $k^h_1>\kappa > k^h_2$ or $k^h_2>\kappa >
k^h_1$, whereas the equations for the 1-strings lead to an integral
equation for the singlet shift function $F_S(k)$, which has the solution
$$\putequation{Fs}$$
The leads to the following result for the singlet phase-shift
$$\delta_S(k^h_1, k^h_2) = 2\pi F_S(k^h_1)\biggr|_{k^h_1>k^h_2}  =
{\pi\over 2}\ .$$
Our result for the complete dressed S-matrix is thus
$$\putequation{Shs}$$
This result ought to be compared with the exact S-matrix for the
nearest neighbour Heisenberg model\upref ft1,ft2/
$$\putequation{sxxx}$$
where $P$ is the $4\times 4$ permutation matrix and
$\mu=\lambda_1-\lambda_2$ is the difference of the spectral parameters
of the two quasiparticles.
We see that (\putlab{Shs}) is the $\mu\to\infty$ limit of the XXX
S-matrix. Note that this does not imply (and it is not true either)
that the low-energy spinon-spinon scattering is the same in HS and XXX
models: the low-energy region of the XXX chain is defined by taking
$\lambda_j\rightarrow\pm\infty, j=1,2$, which still leaves the difference
$\mu$ as a free parameter, the HS chain corresponds to the
$\mu\rightarrow\infty$ limit of the XXX low-enery physics. This fact
does not contradict the identification of the conformal limits of both
XXX and HS chains with the $SU(2)_1$ WZWN CFT. Whereas the conformal
limit of the XXX S-matrix precisely coincides with the WZWN result of
[\putref{zz}] (with trivial LR scattering), the HS result
(\putlab{Shs}) corresponds to ``soft'' scattering in the WZWN model
only (the conformal momenta are taken to be very small).

Like in the case of the XXX antiferromagnet the result (\putlab{Shs})
is a priori exact up to a possible overall constant factor,
which stems from the fact that there are no one-spinon states
and we thus cannot determine the one-particle phase shift directly.

\vskip .5cm\noindent

The form (\putlab{Shs}) of the dressed S-matrix indicates that the
spinons bahave like an ideal gas as the S-matrix is both momentum
independent and proportional to the identity. Furthermore, if we
believe that there is no additional constant phase factor for the
S-matrix, the phase $i$ can be interpreted as exhibiting the
{\sl semionic} character of the spinons as it is {\sl in between} the
phase shifts for an ideal Bose and Fermi gas.

{}From the above discussion the following relation between $SU(2)$ HS
chain and XXX model emerges: the quasiparticles in both models are
spin $\half$ {\sl spinons}, and the $SU(2)$ representation content is
identical in both models. The difference is that the spinons in the HS
model are {\sl noninteracting}, whereas the spinons in the XXX chain
are {\sl interacting}. This follows directly from the form of the
S-matrices and agrees with the picture previously put forward by
Haldane.

{\sc\section{$SU(N)$ Haldane-Shastry Chain}}
\noindent
The $SU(N)$ case can be dealt with in an analogous manner: the ``Bethe
equations'' given by Haldane and Ha\upref hh/ are again very similar
to the ones of the nearest-neighbour $SU(N)$ Sutherland model\upref s3/.
The quasiparticle interpretation\upref ef/ and exact S-matrix\upref
j1,fab/ for the $SU(N)$ Sutherland model have recently been derived
and can be used to analyze the $SU(N)$ HS chain (the important
equations determining the ranges of integers are identical). For
simplicity we we only discuss the $SU(3)$ case, the general case an be
treated analogously. The Ha-Haldane equations for the $SU(3)$ case read
$$\putequation{su3}$$
The ground state is obtained by choosing $M^{(1)}_1={2N\over 3}$ and
$M^{(2)}_1={N\over 3}$ and filling all vacancies for the integers
$I^{(1)1}_\alpha$ and $I^{(2)1}_\alpha$, which corresponds to filling
two ``Fermi seas'' of spectral parameters $k^{(1)1}_\alpha$ and
$k^{(2)1}_\alpha$ between $-\pi$ and $\pi$.
In the thermodynamic limit
we can describe the ground state by densities of spectral parameters
$\rho^{(1)}_1(k)$ and $\rho^{(2)}_1(k)$ (like we did for the $SU(2)$
case above) subject to a set of two coupled integral equations. The
solution of these integral eqautions is straightforward as the
integral kernels are again $\delta$-functions. We find that both
densities are constant over the interval $[-\pi ,\pi]$
$$\putequation{dens}$$
This extends straightforwardly to the general $SU(N)$ case.
Excitations over this ground state can be constructed in an analogous
way to the $SU(2)$-case above. One finds that the only ``dynamical''
objects are {\it holes} in the two Fermi seas of spectral parameters
$k^{(1)1}$ and $k^{(2)1}$, {\it i.e.} only these holes carry energy
and momentum, whereas longer strings (described by spectral parameters
$k^{(j)n},\ n>1,j=1,2$) contribute to neither energy nor momentum
and are only counting degeneracies. All excited states (for the
$SU(N)$ case) can be interpreted as scattering states of $N-1$ types
quasiparticles subeject to {\it superselection rules}. The
quasiparticles are associated with a hole in one of the $N-1$ Fermi
seas respectively, and transform in the $N-1$ fundamental
representations of $SU(N)$. The $SU(N)$-structure of the excited
states as well as the superselction rules are the same as in the
$SU(N)$ Sutherland model, we refer to [\putref{fab}] for a detailed
discussion with proofs of our assertions.

In the $SU(3)$-case there are a total of six quasiparticles, three of
which form the fundamental representations $3$ and $\bar 3$
respectively. The quasiparticles in $3$ have energy and momentum
$$ \epsilon_3(k) = {1\over 3}(\pi^2 - k^2),\qquad p_3(k) = {2\over 3}(\pi
-k)\ ,$$
whereas the quasiparticles in $\bar 3$ have energy and momentum
$$ \epsilon_{\bar 3}(k) = {1\over 6}(\pi^2 - k^2),\qquad p_{\bar 3}(k) =
{1\over 3}(\pi -k)\ .$$

The superselection rules are that the number of quasiparticles of type
$3$ plus twice the number of quasiparticles of type $\bar 3$ must be a
multiple of (the integer number) $3$. That means that the only
two-particle states are given by $3\otimes {\bar 3} = 8\oplus 1$. In
the 3-particle sector only the states $3\otimes 3\otimes 3$ and ${\bar
3}\otimes{\bar 3}\otimes{\bar 3}$ are allowed. It can be shown that
{\sl all} excited states are scattering states of quasiparticles
subject to the superselection rules. Let us now turn to the evaluation
of the phase shifts for the octet and singlet states in the
two-quasiparticle sector. The octet is characterized by choosing
$M^{(1)}_1={2N\over 3}-1$ and $M^{(2)}_1={N\over 3}-1$, which leads to
one hole in the sea of $k^{(1)1}$'s and $k^{(2)1}$ with spectral
parameters $k^{(1)}_h$ and $k^{(2)}_h$ respectively. The Ha-Haldane
equations read $$\putequation{oct}$$ Now, like for the case of the
Hubbard model\upref fk1,fk2/, we have to deal with {\sl two} shift
functions $F_j(k^{(j)1}_\alpha)={{\tilde k}^{(j)1}_\alpha -
k^{(j)1}_\alpha\over k^{(j)}1_{\alpha +1}-k^{(j)}1_\alpha}$, $j=1,2$,
which in the thermodynamic limit are found to obey a system of two
coupled integral equations. The solution of these integral equations
is elementary due to the occurrence of $\delta$-function integral
kernels: $$ F_1(k)={1\over 6}\left( {\rm\ sgn}(k-k^{(2)}_h)-{\rm\
sgn}(k-k^{(1)}_h)\right)= -F_2(k)\ .$$ The dispersion for the octet
states is found to be $E=\epsilon_3(k^{(1)}_h) + \epsilon_{\bar
3}(k^{(2)}_h)$, $P=p_3(k^{(1)}_h) + p_{\bar 3}(k^{(2)}_h)$, in
accordance with the quasiparticle interpretation. The octet
phase-shift is $$\delta_8 = -2\pi F_1(k^{(1)}_h) + 2\pi F_2(k^{(1)}_h)
+ \pi = {\pi\over 3}\ ,$$ where we have use that $k^{(1)}_h>k^{(2)}_h$
for scattering to occur.  The $SU(3)$-singlet in $3\otimes{\bar 3}$ is
obtained by choosing $M^{(1)}_1 = {2N\over 3}-2$, $M^{(1)}_2 = 1$,
$M^{(2)}_1 = {N\over 3}-2$ and $M^{(2)}_2 = 1$. Energy and momentum of
the singlet are the same as for the octet. The shift-functions can be
determined analogously to the octet case, although the computation is
slightly more difficult due to the presence of the 2-strings. The
result is $$\delta_1 = {\pi\over 3}=\delta_8\ ,$$ {\sl i.e.} the
singlet phase shift is the same (constant) as the octet phase shift.
The phase-shifts for scattering of quasiparticles of type $3$ ($\bar
3$) on quasiparticles of type $3$ ($\bar 3$) can be extracted from the
three-particle states $3\otimes 3\otimes 3$ (${\bar 3}\otimes {\bar
3}\otimes {\bar 3}$), with the result that the phase-shifts for
scattering of $3$ on $3$ and $\bar 3$ on $\bar 3$ are also equal to
$\pi\over 3$. This implies that the quasiparticles are an ideal gas
with fractional statistics ``$\pi\over 3$''. For the $SU(N)$ case we
conjecture the phase to be $\pi\over N$. The interesting new
phenomenon is the ``decoloration'' of physical excitations: the
superselection rules force the quasiparticles to combine to either
``mesons'' ($3\otimes{\bar 3}$) or ``baryons''
($3\otimes 3\otimes 3$ and ${\bar 3}\otimes {\bar 3}\otimes {\bar
3}$). In this way the $SU(3)$ HS chain (as well as the $SU(3)$
Sutherland model\upref fab/) reminds an ideal (1-d) gas of (confined)
quarks. Like for the $SU(2)$ case, the constant S-matrix found for the
HS model is precisely the limit $\mu\rightarrow\infty$ of the
corresponding S-matrix of the nearest neighbour Sutherland model\upref
fab/, where $\mu$ is the difference of the spectral parameters of the
two scattering quasiparticles.

{\sc\section{ Calogero-Sutherland Model}} \noindent The
Calogero-Sutherland model\upref s1,s2,cal/ is given by the following
hamiltonian
$$\putequation{Hcs}$$
Ground state, excitations, and
thermodynamics for the CS model are all constructed from the following
set of ``asymptotic''Bethe equations\upref s2/
$$\putequation{baecs}$$
where $S(k)=-\exp(-i\pi (\lambda -1) {\rm sgn}(k))$ is the {\sl bare}
S-matrix describing scattering of two {\it bare} particles over the
bare (trivial) vacuum. Following the logic of Bethe Ansatz, Sutherland
used (\putlab{baecs}) to construct the true ground state and {\sl
dressed} excitations over it, by ``filling the Fermi sea''. By this we
mean the following: taking the logarithm of (\putlab{baecs}) we arrive
at the set of equations $$\putequation{cslog}$$ where $I_j$ are all
integer of half-odd integer numbers, which can be chosen as a set of
quantum numbers that completely determines an eigenstate of the
hamiltonian. The ground state is characterized by filling all
vacancies for for the $k_j$'s symmetrically around zero in the
interval $[-k_F,k_F]$. The ``Fermi-momentum'' is $k_F=\sqrt{\mu}$,
where $\mu$ is the chemical potential. In analogy to the
delta-function Bose gas\upref lieb,liebl/ {\it dressed} (particle-hole)
excitations can be constructed by removing one particle with rapidity
$k_h$ from the Fermi-sea and placing it on a vacancy $k_p$ outside the
Fermi sea. Energy and momentum of a particle-hole excitation are given by
$E_{ph}=(k_p^2-\mu) + ({\mu-k_h^2\over\lambda})$,
$P_{ph}=(k_p-k_F)+({k_F-k_h\over\lambda})$.  Equations
(\putlab{cslog}) can also be used to determine the phase shifts for
scattering of dressed excitations over the ground state. The
computations are completely analogous to the ones for the
$\delta$-function Bose gas [\putref{vek2,vek3}], so that we will only
give the results here.  It is again straighforward to show that all
phase shifts are momentum-independent constants, which proves the
ideal-gas nature of the quasiparticles. In determining the constants
we follow [\putref{vek2}] (where the $\delta$-function Bose gas was
treated, see p.23) and change the boundary conditions for one-particle
and one-hole excitations (the situation here is quite analogous to the
Bose gas case).  We then find that particles do not get any dressing
through the ground state, {\it i.e.} still behave like bare particles,
which scatter off each other with the bare phase-shift
$\delta_{pp}=-\pi\lambda$. This is in agreement with the
results previously obtained in [\putref{p1}] by means of different
methods. Particles scatter off holes with phase-shift
$\delta_{ph}=\pi\lambda$, which means that to the scattering particle
a hole is nothing by the absence of a bare particle. Last but not
least the hole-hole phase-shift is $\delta_{hh}={\pi\over\lambda}$.

We not that we recover the correct scattering phases for free fermions
$e^{\delta_{pp}}=e^{\delta_{hh}}=-1$ for $\lambda=1$ and free bosons
$e^{\delta_{pp}}=0$ for $\lambda=0$ (in this limit there are no holes
but only particles).

{\sc\section{Other ${1\over\sin^2(r)}$-Type Models}}
\noindent
Other candidates for applying Korepin's dressed S-matrix method would
be for example the $gl(n,1)$ supersymmetric ``$t$-$J$'' models with
long-range exchange interactions\upref ky,k1,k2/, or Kawakami's
hierarchy of $SU(N)$ electron models\upref k3/. The asymptotic Bethe
equations (ABE)\upref k1,k2,k3/ have to be complemented by a
``squeezed-string'' prescription\upref hh/ in order to give the
correct degeneracies of the spectrum. If we take into account only the
states given by the ABE we find that all phase-shifts will be
constants, and the states described by the ABE will thus describe
mixtures of ideal gases with fractional statistics.
This can be seen as follows: from the computations above it is clear
that the ideal-gas character of the quasiparticles is ``caused'' by
the $\delta$-function kernel in the integral equations, or
alternatively the $\sgn(x)$-kernels in the Asymptotic Bethe equations.
The ocurrence of $sgn(x)$-kernels in the ABA equations is generic
feature of models with ${1\over\sin^2(r)}$-type interactions, so
that by analyzing only ABA states we would conclude that all these
models describe mixtures of noninteracting quasiparticles.

However, there are a number of open problems concerning the
supersymmetric models: the squeezed-string prescription seems to be
not available, but more importantly, the ground state will in general
{\sl not} be a $gl(n,1)$ singlet and thus no $Y(gl(n,1))$ Yangian
singlet. This is easily seen for the case of the long-range
supersymmetric ($gl(2,1)$) $t$-$J$ model: in the $3^L$-dimensional
Hilbert space without doubly occupied sites there exists no $gl(2|1)$
singlet! Thus, very much like for the case of the nearest neighbour
model\upref fk/, the ground state will belong to a larger $gl(2,1)$
multiplet. This raises the question how to interpret the other states
in the multiplet containing the ground state in terms of a quasiparticle
picture, which is necessary for identifying the model as a gas.

{\sc\section{Interacting Quasiparticles: ${1\over\sinh^2(r)}$-Models}}
\noindent
Let us now demonstrate that not all models with long-range
interactions describe noninteracting quasiparticles. To this end let us
consider the ${1\over\sinh^2(r)}$-CS model\upref s4/, defined in
terms of the hamiltonian ($\lambda\geq 1$)
$$\putequation{Hcsh}$$
In the limit $a\rightarrow\infty$ (\putlab{Hcsh}) reduces to the CS
model with coupling $2\lambda(\lambda-1)a^2$.
The ABA as well as ground state and excitations were constructed in
[\putref{s4}]. The ABA equations are
$$\putequation{cshlog}$$
where $a$ has been set to $1$ and where
$$\putequation{cshkern}$$
In terms of the variables $k_j$ the effect of $a$ is recovered by a
rescaling $k_j\rightarrow a k_j$.
Excitations over the ground state are (like in the CS case above) of
particle-hole type. Their energy and momentum are
$E_{ph}=\epsilon(k_p) - \epsilon(k_h)$ and $P_{ph} = p_{k_p}-p_{k_h}$,
where $k_p$ and $k_h$ are the rapidities of the particle and hole
respectively, and where $\epsilon(k)$ and $p(k)$ are given in
terms of the integral equations
$$\eqalign{\epsilon(k) &= k^2-\mu -{1\over 2\pi}\int_{-B}^B dk^\prime\
\epsilon(k^\prime) \theta^\prime(k-k^\prime)\ ,\cr
p(k) &= k -{1\over 2\pi}\int_{-B}^B dk^\prime\ p(k^\prime)
\theta^\prime(k-k^\prime)\ ,\cr }$$
where $\theta^\prime(x) = {d\over dx} \theta(x)$.
Here $\mu$ is the chemical potential, and the integral boundary $B$ is
determined as a function of $\mu$ through the requirement that
$\epsilon(\pm B)=0$.
The computations of the S-matrices for scattering of particles on
holes, particles on particles, and holes on holes are again completely
analogous to the ones for the Bose gas (see p.23 of [\putref{vek2}]).
The result is that all phase-shifts can be expressed in terms of a
function $\delta(\lambda,\mu)$ subject to the integral equation
$$\putequation{cshshift}$$
The S-matrix for particle-hole scattering ($k_p>B, -B\leq k_h\leq B$,
the constraint $k_p>B$ is only a matter of convenience) is given as
$$S_{ph}(k_p,k_h) = e^{i\delta(k_p,k_h)}\ .$$
Similarly particle-particle and hole-hole S-matrices are found to be
$$\eqalign{S_{pp}(k_{p,2},k_{p,1}) &= e^{-i\delta(k_{p,2},k_{p,1})}\
,\quad k_{p,2}>k_{p,1}>B \cr
S_{hh}(k_{h,2},k_{h,1}) &= e^{-i\delta(k_{h,2},k_{h,1})}\
,\quad B\geq k_{h,2}>k_{h,1}\geq -B\  .\cr }$$

As noted avove the CS limit is obtained by rescaling $k_j\rightarrow a
k_j$ and then taking $a\rightarrow\infty$ keeping
$2\lambda(\lambda-1)a^2=: 2g(g-1)$ fixed. In this limit one obtains
$\theta(k_1-k_2)\rightarrow \pi(g-1)\sgn(k_1-k_2)$\upref s4/,
and the expression for the S-matrices reduce to the ones found for the
CS model in section $4$ (if $g$ is identified with $\lambda$ of
section $4$) as can be seen directly from (\putlab{cshshift}).
In general the integral equation (\putlab{cshshift}) can only be
solved numerically, the result being a nontrivial function of $k_1$
and $k_2$.

Physically our results for the S-matrices imply that the
quasiparticles in the ${1\over\sinh^2(r)}$-CS model are {\sl
interacting} as the S-matrices are momentum dependent and nontrivial.
In the ${1\over r^2}$-limit they become {\sl noninteracting}.

A very interesting extension of the $1\over \sinh^2(r)$-CS model is
the $1\over \sinh^2(r)$-CS model with exchange\upref p3,srs,ino/. The
hamiltonian of the model is\upref ss/
$$\putequation{Hcshx}$$
where $P_{jk}$ is a permutation operator exchanging the spins of the
particles at positions $x_j$ and $x_k$. We will consider only the
simplest case of $SU(2)$ spins. $N$ is the number of particles in a
box of length $L$ and we are interested in the limit
$L\rightarrow\infty$ keeping the density ${N\over L}$ fixed.
In the inverse square limit $\kappa\rightarrow 0$ the interaction
becomes $\sum_{j<k}{\lambda(\lambda- P_{jk})\over (x_j-x_k)^2 }$ and
the model reduces to the CS model with inverse square exchange\upref
p3,ss,bghp, k4/. The ABA equations for (\putlab{Hcshx}) are \upref srs/
$$\putequation{cshaba}$$
Here $\kappa$ has been set to $1$ and the effect of $\kappa$
corresponds to a rescaling $k_j\rightarrow {k_j\over\kappa}$ and
$\alpha_s\rightarrow {\alpha_2\over\kappa}$.
All $k_j$'s are real (complex $k$'s do not lead to bound states in the
bare scattering amplitudes on the r.h.s. of the first equation in
(\putlab{cshaba})), whereas the $\alpha_s$'s can form bound states of the
form $\alpha_s^{n,j} = \alpha_s^n + i (n+1-2j)\lambda$ with
$\alpha_s^n\in\Real$. This not surprising as the second set of
equations in (\putlab{cshaba}) is nothing but the set of Bethe
equations for an inhomogeneous Heisenberg model. Inserting this
``string hypothesis'' into (\putlab{cshaba}) and then taking the
logarithm we obtain
$$\putequation{cshlogx}$$
where $I_j$ and $J^n_\alpha$ are integer or half-odd integer numbers,
$\theta(x)$ is given by (\putlab{cshkern}), $\vartheta(x) =
2\arctan({x})$ and where
$${\vartheta_{nm}(x) = \cases{\vartheta ({x\over{|n-m|}}) +
2\ \vartheta ({x\over{|n-m|+2}})+\dots +2\ \vartheta ({x\over{n+m-2}}) +
\vartheta ({x\over{n+m}})&if $n\ne m$\cr\cr
2\ \vartheta ({x\over{2}}) + 2\ \vartheta ({x\over{4}})+\dots +
2\ \vartheta ({x\over{2n-2}})+\vartheta ({x\over{2n}})&if $n=m$\ \
.\cr}}$$
The range of $J^n_\alpha$ follows from (\putlab{cshlogx}) to be
$$\putequation{intx}$$
where $M_m$ is the number of $\alpha$-strings of length $m$.

The construction of ground state and excitations is rather similar to
the less than half-filled Hubbard model\upref coll/.
The ground state is obtained by filling two Fermi seas of spectral
parameters $k_j$ and $\alpha^1_s$. In the thermodynamic limit it is
described in terms of two densities (of spectral parameters) $\rho(k)$
and $\sigma(\alpha)$ subject to the coupled integral equations
$$\eqalign{\rho(k)&={1\over 2\pi} - {1\over 2\pi}\int_{-A}^A dk^\prime\
\theta^\prime(k-k^\prime)\ \rho(k^\prime) + {1\over 2\pi}
\int_{-\infty}^\infty d\alpha\ {2\lambda\over\lambda^2+(k-\alpha)^2}
\sigma(\alpha)\ ,\cr
\sigma(\alpha)&={1\over 2\pi}\int_{-A}^A dk\
{2\lambda\over\lambda^2+(k-\alpha)^2}\ \rho(k) -  {1\over
2\pi}\int_{-\infty}^\infty d\alpha^\prime\
{4\lambda\over 4\lambda^2+(\alpha-\alpha^\prime)^2}\
\sigma(\alpha^\prime)\ ,\cr  }$$
where $\int_{-A}^A dk\ \rho(k) = {N\over L}$ and where the integration
boundary $A$ is a function of the chemical potential $\mu$.
The ground state energy density is given by $E_{GS} =
\int_{-A}^A dk\ \rho(k) k^2$ ($\mu = {dE_{GS}\over dN}$, which fixes
$A$ as a function of $\mu$). We note that for $\kappa >0$,
$\sigma(\alpha)>0$ on the whole real axis (``the sea of $\alpha$'s is
completely filled''), whereas in the inverse square limit
$\kappa\rightarrow 0$, $\sigma(\alpha)=0\ \forall|\alpha|>A$. This is
in agreement with [\putref{k3}]. There are two classes of
low-lying excitations over the ground state: particle-hole excitations
in the Fermi sea of $k$'s, which are very similar to the excitations
in the $1\over\sinh^2(r)$ CS model (see above; the only
difference is that now there will be a dressing through the second
Fermi sea of $\lambda$'s) and spin-excitations in the second Fermi sea.
We will constrain ourselves to a discussion of the spin-excitations
here.
Inspection of (\putlab{intx}) shows that the situation for $\kappa>0$
is very similar to the one for the HS chain treated in section $2$:
the simplest low-lying excitations are a spin-triplet ($M_1 =
{N\over2}-1$) and a spin-singlet ($M_1={N\over 2}-2, M_2=1$)
two-hole excitation. In the inverse square limit the Fermi sea of
$\alpha$'s is not completely filled, so that the simplest spin
excitations are of particle-hole type. We consider only the case
$\kappa>0$ as it is the far more interesting one.
Like in sections $2$ and $3$ we describe the excitations in terms of
shift-functions $F_1(k)$ and $F_2(\alpha)$ (the construction is very
similar to the one of spin excitations in the Hubbard model, which was
treated in detail in [\putref{fk1}]). After some manipulations we find
for the triplet
$$\putequation{Ftx}$$
where $\alpha_{h,p}$ are the rapidities of the two holes and where
$$\Theta(x) = {\rm Re}\left\{{1\over 2\lambda}\Psi(\half +i{x\over
4\lambda}) - {1\over 2\lambda}\Psi(1 +i{x\over 4\lambda})
+\Psi(\lambda+i{x\over 2}) - \Psi(1+i{x\over 2})\right\} .$$
Here $\Psi(x)$ is the Digamma function. Energy and momentum of the
spin-triplet are given by
$$\eqalign{E_{ST}(\alpha_{h,1},\alpha_{h,2})&=\int_{-A}^Adk\
2k\ F^T_1(k)\ ,\cr
P_{ST}(\alpha_{h,1},\alpha_{h,2})&=\int_{-A}^Adk\ F^T_1(k)\ .\cr}$$
The scattering phase shift is given as $\delta_T(\alpha_{h,1},
\alpha_{h,2}) = 2\pi F^T_2(\alpha_{h,1})$ with $\alpha_{h,1} -
\alpha_{h,2}>0$. For the spin-singlet we find
$$\putequation{Fsx}$$
where $F^T_{1,2}$ are given by (\putlab{Ftx}). From the first equality
in (\putlab{Fsx}) it follows immediately that energy and momentum of
triplet and singlet are identical (as it must be). The singlet
scattering phase-shift is found to be
$\delta_S(\alpha_{h,1}, \alpha_{h,2}) = 2\pi F^S_2(\alpha_{h,1})$.
The resulting two particle S-matrix describing scattering of spinons
in the $1\over\sinh^2(r)$ CS model with exchange is
$$\putequation{smx}$$
where $P$ is the $4\times 4$ permutation matrix and where
$F^T_2(\alpha)$ is given by (\putlab{Ftx}). The result we get is
extremely similar to the spinon-spinon S-matrix of the Hubbard model\upref
fk1,fk2, natan/: the rapidities get renormalized by a factor of $2\lambda$
($2U$ in the Hubbard model) as compared to the pure XXX scattering
matrix (\putlab{sxxx}), and the common overall phase gets an additonal
contribution from the dynamical degrees of freedom ({\sl i.e.} the
Fermi sea of $k$'s). We can rewrite (\putlab{smx})
in terms of the XXX S-matrix $S_{XXX}$ given by (\putlab{sxxx}) as
$$S(\alpha_{h,1}, \alpha_{h,2}) = S_{XXX}(\nu)\
\exp\left({{2\pi i\over 4\lambda}\int_{-A}^A\!\!dk {F^T_1(k)\over
\cosh\left({\pi\over 2\lambda} (\alpha_{h,1}-k)\right)}}\right)\ ,$$
where $\nu$ is as in (\putlab{smx}). In the limit
$\nu\rightarrow\infty$ this reduces to the Haldane-Shastry result
(\putlab{Shs}). We see that in (\putlab{smx})
there are two two distict contributions: one from pure spin-spin
scattering (given by $S_{XXX}(\nu)$), and one from coupling of spin
and dynamical degrees of freedom (given by the second factor).

As was noted by Sutherland, R\"omer and Shastry, it is possible to
freeze out the dynamical degrees of freedom in (\putlab{Hcshx}) by
taking the limit $\lambda\rightarrow\infty$\upref srs/. In this limit
the particles freeze into an equidistant lattice $x_j={j\over d}$
(recall that $d={N\over L}$ is the fixed density of particles), and
the hamiltonian (\putlab{Hcshx}) separates into  $H_{dyn}+2\lambda
H_{latt}$, where $H_{dyn}$ is of the form (\putlab{Hcsh}) with coupling
$\lambda(\lambda-1)$ and where
$$\putequation{Hlatt}$$
Ground state and excitations of the lattice model (\putlab{Hlatt}) can
be obtained by rescaling and expanding the spectral parameters in
(\putlab{cshlogx}) according to $k_j=2\lambda \xi^{(0)}_j+\xi^{(1)}_j+
{1\over 2\lambda}\xi^{(2)}_j+ ...$, $\alpha^n_s=2\lambda \beta^n_s+ ...$,
and then expanding the ABE in inverse powers of $\lambda$. This
procedure yields
$$\putequation{baespin}$$
where $\theta_0(x)={x\over 2}\ln\left(1+{1\over
x^2}\right)+{i\over2}\ln\left({1-ix\over 1+ix}\right)$ and $\theta_1(x)=
-{\pi\over 2}-{i\over 2} \ln\left({1-ix\over 1+ix}\right)$ .
The first set of equations (\putlab{baespin}) is of order $\lambda$,
and leads in the
thermodynamic limit to an integral equation for the ground state density
of the dynamical part\upref srs/ $\rho(x)$ (defined to be the limit
$N\rightarrow\infty$ of $\rho(\xi^{(0)}_j)=
{1\over N(\xi^{(0)}_{j+1}-\xi^{(0)}_j)}$
The second and third sets of equations (\putlab{baespin})
are of order $1$ and can be used to construct ground state and excitations
of the spin model (\putlab{Hlatt}). This has already been done (for the
general $SU(N)$) case in [\putref{srs}].
Our goal here is to determine the exact S-matrix for the $SU(2)$ case,
for which we need to construct all two particle excitations in the
framework of the $F$-function formalism. This is easily done as the
integers $J^n_s$ are actually the same as in the dynamical model treated
above. Before we get to this let us review some results of [\putref{srs}]
that we will need later on. The ground state of (\putlab{Hcshx}) in the limit
$\lambda\rightarrow\infty$ is obtained by taking $M_1={N\over 2},
\ M_k=0\ \forall k>1$. In the thermodynamic limit $N\rightarrow\infty$
($d={N\over L}$ fixed) the ABA
equations turn into a set of three coupled integral equations\upref srs/
$$\putequation{srsgs}$$
where $a$ is a function of the fixed density $d$ and
where $\sigma(\beta)$ and $\gamma(\xi)$ are the infinite volume limits of
the densities ${1\over N(\beta^1_{s+1}-\beta^1_s)}$ and ${\xi_j^{(1)}\over
N(\xi^{(0)}_{j+1}-\xi^{(0)}_j)}$.
In order to describe only the ground state of (\putlab{Hlatt})
it is necessary to decouple the dynamical degrees of freedom by
hand\upref srs/. We note that for the excitations no such decoupling
has to be carried out because the structure of (\putlab{srsgs}) is
such that the dynamical degrees of freedom decouple automatically.
Eqns (\putlab{srsgs}) and (\putlab{baespin}) are all we need to
determine the S-matrix. Let us start with the spin-triplet phase shift.
The spin-triplet excitation is obtained by taking $M_1={N\over 2}-1$
and all other $M_k=0$. There are two holes with corresponding spectral
parameters $\beta_{h,j},\ j=1,2$ in the distribution of $\beta$'s.
The ABA equations (\putlab{baespin}) read (our convention is
${\tilde J}_s -J_s={1\over 2}$)
$$\putequation{baespinT}$$
Subtracting the corresponding ground state equations
from (\putlab{baespinT}) we
obtain coupled equations for the shift functions $F_2(\beta_j)={
{\tilde\beta}_j-\beta_j\over\beta_{j+1}-\beta_j}$ and
$F_1(\xi_j)={{\tilde \xi}^{(1)}_j-\xi_j^{(1)}\over
\xi^{(0)}_{j+1}-\xi^{(0)}_j}$, which in the thermodynamic limit turn
into coupled integral equations
$$\putequation{ftsinh}$$
Note that in order to obtain (\putlab{ftsinh}) we used the ground state
equations (\putlab{srsgs}). The equation for $F_2$ is readily solved
by Fourier techniques
$$F_2(\beta)= {1\over 2}-{i\over 2\pi}\sum_{p=1}^2 \ln\left({
\Gamma\left({1+i{\beta-\beta_{h,p}\over 2}\over 2}\right)
\Gamma\left(1-i{\beta-\beta_{h,p}\over 2}\right)\over
\Gamma\left({1-i{\beta-\beta_{h,p}\over 2}\over 2}\right)
\Gamma\left(1-i{\beta-\beta_{h,p}\over 2}\right)}\right) .$$
The triplet phase shift is $\delta_T=2\pi F_2(\beta_{h,1})$ with
$\beta_{h,1}-\beta_{h,2}>0$ and is {\sl identical} to the triplet phase
shift in the nearest neighbour XXX model !
The excitation energy of the triplet states is
$$\eqalign{E_T&=\int_{-a}^a d\xi\ 2\xi\ F_1(\xi)= -\sum_{j=1}^2
\int_{-a}^a d\xi\ {e(\xi)\over 2\cosh\left(\pi(\xi-\beta_{h,j})\right)}\cr
\xi^2-\mu&=\int_{-a}^a d\xi^\prime\ e(\xi^\prime)
\theta_0^\prime(\xi-\xi^\prime)\ , \cr}$$
where $e(\xi)$ is the "classical ground state energy density" of Sutherland,
R\"omer and Shastry\upref srs/. Repeating the above steps for the spin
singlet ($M_1={N\over 2}-2$, $M_2=1$) we find that the excitation energy
is the same as for the triplet, and the phase-shift is
$\delta_S=\delta_T+2\arctan\left({\beta_{h,1}-\beta_{h,2}}\right)
- \pi$, which results in an S-matrix identical to the nearest
neighbour Heisenberg XXX S-matrix (\putlab{sxxx}) with $\mu=\beta_{h,1}-
\beta_{h,2}$. This shows that the spinons in the nearest neighbour
Heisenberg model and its ${1\over\sinh^2(r)}$ analog are very similar:
in both models they are interacting with the {\sl same} S-matrix, the
only difference is the dispersion. Our result for the S-matrix furthermore
leads to the conclusion that the conformal limit of the
${1\over\sinh^2(r)}$ model (\putlab{Hlatt}) is given by the $SU(2)_1$
WZWN conformal field theory.

{\sc\section{Discussion}}
In this note we have determined the dressed scattering matrices for
several models with long-range interactions by applying a method
invented by Korepin for models solvable by (normal) Bethe Ansatz.
We would like to stress that this method can be applied to {\it any}
model, for which the Asymptotic Bethe Ansatz can be formulated.
Our results show very directly that models with $1\over\sin^2(r)$
interaction are ideal gases with fractional statistics. Long-range
models with $1\over\sinh^2(r)$ interactions describe {\sl interacting}
elementary excitations and are close in nature to integrable
nearest-neighbour models. Our analysis in section $7$ can readily be
generalized from $SU(2)$ to $SU(N)$. The structure of the ABE relevant
for the spin degrees of freedom is that of an inhomogeneous $SU(N)$
Sutherland model\upref s3/. On the basis of our results for $SU(2)$
we conjecture that the resulting dressed
S-matrix for the $SU(N)$ spin chain with ${1\over\sinh^2(r)}$ hopping
is identical to the one for the nearest neighbour model.
The fact that elementary excitations in ${1\over\sinh^2(r)}$ models
are interacting in basically the same way as in their nearest neighbour
analogs indicates that the evaluation of correlation functions may be
rather more difficult than for the ${1\over\sin^2(r)}$ case, in which
elementary excitations are free.

Finally we would like to point out a close relation between
fractional statistics and the fractional ``charge'' previously observed
in many solvable models. As was first observed by Korepin for the case
of the Massive Thirring model (MTM)\upref vek1/, elementary
excitations over the true ground state will in general carry a
fractional charge. Here charge is the eigenvaule of the fermion number
operator defined in terms of the (fermionic) quantum fields entering
the hamiltonian. The relation to fractional statistics is most easily
seen for the simple example of $SU(2)$ XXX model: the analog of charge
is the third component of the spin. A one-particle excitation over the
bare vacuum corresponds to flipping one spin, and thus carries
``charge'', {\sl i.e.} spin $1$. By construction this excitation has
bosonic statistics. From the discussion above we see that flipping one
spin over the true (antiferromagnetic) ground state leads to a {\sl
two}-spinon excitation, and that one spinon thus carries ``charge''
$1\over 2$, and carries fractional statistics. Analogously we can
deduce that the quasiparticles with fractional charge in the MTM ought
to be thought of as objects of fractional statistics as well.

\vskip .5cm
\centerline{\sc Acknowledgements:}
\vskip .5cm
I am grateful to the Erwin Schr\"odinger International Institute for
Mathematical Physics in Vienna, where most of this work was done,
for hospitality. It is a pleasure to thank H.-P. Eckle, H. Grosse,
V.E. Korepin, V. Pasquier and V. Rittenberg for stimulating
discussions, and R.A. R\"omer and B. Sutherland for helpful
correspondence.
\sectionnumstyle{Roman}
\sectionnum=0
\vfil
\begin{putreferences}
\centerline{{\sc References}}
\vskip .5cm
\reference{pas}{F. Lesage, V. Pasquier, D. Serban,\ {\sl cond-mat
    9405008\ .}}
\reference{ha}{Z.N.C. Ha,\ {\sl cond-mat 9405063\ .}}
\reference{HaBe}{H. Bethe,\ \ZP{79}{1931}{205}.}
\reference{mt}{M. Takahashi,\ \PTP{46}{1971}{401}.}
\reference{h1}{F.D.M. Haldane,\ in {\sl Correlation Effects in
Low-Dimensional Electron Systems}, Eds. A. Okiji and N. Kawakami,
Springer Series in Solid-State Sciences 118, Springer (1994).}
\reference{h2}{F.D.M. Haldane,\ \PRL{60}{1988}{635}.}
\reference{h3}{F.D.M. Haldane,\ \PRL{67}{1991}{937}.}
\reference{h4}{F.D.M. Haldane,\ \PRL{66}{1991}{1529}.}
\reference{hh}{Z.N.C. Ha, D. Haldane,\ \PRB{46}{1992}{9359}.}
\reference{s1}{B. Sutherland,\ \JMP{12}{1971}{246}.}
\reference{s2}{B. Sutherland,\ \JMP{12}{1971}{251}.}
\reference{s3}{B. Sutherland,\ \PRB{12}{1975}{3795}.}
\reference{s4}{B. Sutherland,\ {\sl Rocky Mountain Journal of
Mathematics} {\bf 8} (1978) 413 .}
\reference{sh}{B.S. Shastry,\ \PRL{60}{1988}{639}.}
\reference{ss}{B. Sutherland, B.S. Shastry,\ \PRL{71}{1993}{5}.}
\reference{srs}{B. Sutherland, R. R\"omer, B.S. Shastry,\
\PRL{73}{1994}{2154}.}
\reference{rs}{B. Sutherland, R. R\"omer,\ \PRL{71}{1993}{2789}.}
\reference{ef}{F.H.L. E\char'31ler, H. Frahm,\ {\sl unpublished}.}
\reference{fab}{F.H.L. E\char'31ler,\ {\sl in preparation}.}
\reference{fk1}{F.H.L. E\char'31ler, V.E. Korepin,\ \NPB{426}{1994}{505}.}
\reference{fk2}{F.H.L. E\char'31ler, V.E. Korepin,\ \PRL{72}{1994}{908}.}
\reference{j1}{H. Johannesson,\ \NPB{270}{1986}{235}.}
\reference{ft1}{L.D. Faddeev, L. Takhtajan, \JSM{24}{1984}{241}.}
\reference{ft2}{L.D. Faddeev, L. Takhtajan,\ \PLA{85}{1981}{375}.}
\reference{vek1}{V.E. Korepin,\ \TMP{41}{1979}{953}.}
\reference{vek2}{V.E. Korepin, A.G. Izergin, N.N. Bogoliubov,\ {\sl
    Quantum Inverse Scattering Method and Correlation Functions},
  Cambridge University Press, 1994.}
\reference{vek3}{V.E. Korepin, G. Izergin and N.M. Bogoliubov,\ {\sl
Exactly Solvable Problems in Condensed Matter and Relativistic field
Theory},\ B.S. Shastry, S.S. Jha, V. Singh (eds.)\ Lecture Notes in
Physics, v.242, Berlin: Springer Verlag, (1985), p.220.}
\reference{al}{N. Andrei, J.H. Lowenstein,\ \PLA{80}{1980}{401}.}
\reference{bw}{D. Bernard, Y.S. Wu,\ {\sl cond-mat 9404025}\ .}
\reference{bghp}{D. Bernard, M. Gaudin, F.D.M. Haldane, V. Pasquier,\
  \JPA{26}{1993}{5219}.}
\reference{barry}{B. McCoy,\ {\sl private communication}.}
\reference{cal}{F. Calogero, \JMP{10}{1969}{2191}.}
\reference{j1}{H. Johannesson,\ \NPB{270}{1986}{235}.}
\reference{ky}{Y. Kuramoto, H. Yokoyama,\ \PRL{67}{1991}{2493}.}
\reference{k1}{N. Kawakami,\ \PRB{46}{1992}{1005}.}
\reference{k2}{N. Kawakami,\ \PRB{45}{1992}{7525}.}
\reference{k3}{N. Kawakami,\ \PRL{71}{1993}{275}.}
\reference{k4}{N. Kawakami,\ in {\sl Correlation Effects in
Low-Dimensional Electron Systems}, Eds. A. Okiji and N. Kawakami,
Springer Series in Solid-State Sciences 118, Springer (1994).}
\reference{fk}{A. F\"orster, M. Karowski,\ \NPB{396}{1993}{611}.}
\reference{p1}{A. Polychronakos,\ \NPB{324}{1989}{597}.}
\reference{p2}{A. Polychronakos, J. Minahan,\ {\sl hep-th 9404192}.}
\reference{p3}{A. Polychronakos,\ \PRL{69}{1992}{703}.}
\reference{coll}{C.F. Coll,\ \PRB{9}{1974}{2150}.}
\reference{natan}{N. Andrei,\ {\sl Integrable Models in Condensed
Matter Physics}, preprint cond-mat 9408101.}
\reference{ino}{V.I. Inozemtsev,\ \JSP{59}{1990}{1143}.}
\reference{liebl}{E.H. Lieb, W. Liniger,\ \PR{130}{1963}{1605}.}
\reference{lieb}{E.H. Lieb,\ \PR{130}{1963}{1616}.}
\reference{zz}{A.B. Zamolodchikov, Al.B. Zamolodchikov,\
\NPB{379}{1992}{602}.}
\end{putreferences}
\bye